\documentclass[pdflatex,sn-basic]{sn-jnl}

\usepackage{graphicx}%
\usepackage{multirow}%
\usepackage{amsmath,amssymb,amsfonts}%
\usepackage{amsthm}%
\usepackage{mathrsfs}%
\usepackage[title]{appendix}%
\usepackage{xcolor}%
\usepackage{textcomp}%
\usepackage{manyfoot}%
\usepackage{booktabs}%
\usepackage{algorithm}%
\usepackage{algorithmicx}%
\usepackage{algpseudocode}%
\usepackage{listings}%
\usepackage{natbib}
\usepackage{amsfonts}

\usepackage{lmodern}
\usepackage[pagewise]{lineno}
\usepackage{multirow}
\usepackage{adjustbox}
\usepackage{enumitem}
\usepackage{comment}
\usepackage{booktabs}
\usepackage{longtable}
\usepackage{array}
\usepackage{ragged2e}
\usepackage{caption}
\usepackage{subcaption}

\newcolumntype{Y}{>{\RaggedRight\arraybackslash}p{5cm}}
\newcolumntype{Z}{>{\RaggedRight\arraybackslash}p{3.8cm}}

\theoremstyle{thmstyleone}%
\theoremstyle{thmstyletwo}%

\theoremstyle{thmstylethree}%

\begin{document}

\begin{titlepage}
    \vspace*{\fill}
    \begin{center}
        \Large
        \textbf{LR-Robot: An Human-in-the-Loop LLM Framework for Systematic Literature Reviews with Applications in  Financial Research
 } \\[1.5cm]

        \large
        Wei Wei\textsuperscript{a}\\
        Jin Zheng\textsuperscript{a}\\
        Zining Wang\textsuperscript{a}\\
        Weibin Feng\textsuperscript{a}\\ [1cm]

        \normalsize
        \textsuperscript{a}School of Engineering Mathematics and Technology, University of Bristol, Ada Lovelace Building, Tankard's Close, Bristol, BS8 1TW, England, United Kingdom \\

        \textbf{Corresponding author:} Jin Zheng \\
        Email: jin.zheng@bristol.ac.uk \\[1cm]
        


    \end{center}
    \vspace*{\fill}

\end{titlepage}

\title{}










\abstract{
The exponential growth of financial research has rendered traditional systematic literature reviews (SLRs) increasingly impractical, as manual screening and narrative synthesis struggle to keep pace with the scale and complexity of modern scholarship. While the existing artificial intelligence (AI) and natural language processing (NLP) approaches often often produce outputs that are efficient but contextually limited, still requiring substantial expert oversight. \\

To address these challenges, we propose LR-Robot, a novel framework in which domain experts define multidimensional classification taxonomies and prompt constraints that encode conceptual boundaries, large language models (LLMs) execute scalable classification across large corpora, and systematic human-in-the-loop evaluation ensures reliability before full-dataset deployment.The framework further leverages retrieval-augmented generation (RAG) to support downstream analyses including temporal evolution tracking and label-enhanced citation networks.\\

We demonstrate the framework on a corpus of 12,666 option pricing articles spanning 50 years, designing a four-dimensional taxonomy and systematically evaluating up to eleven mainstream LLMs across classification tasks of varying complexity. The results reveal the current capabilities of AI in understanding and synthesizing literature, uncover emerging trends, reveal structural research patterns, and highlight core research directions. By accelerating labor-intensive review stages while preserving interpretive accuracy, LR-Robot provides a practical, customizable, and high-quality approach for AI-assisted SLRs.\\

}





\maketitle

\newpage
\section{Introduction}
\label{sec1}

The volume of financial research has expanded exponentially, with thousands of annual publications across asset pricing, risk management, and econometrics. While Systematic Literature Reviews (SLRs) are the cornerstone of rigorous scholarship, the current scale of the literature has rendered traditional, expert-driven methodologies increasingly impractical. The logical bottleneck lies in the manual nature of conventional approaches: relying on individual screening and narrative synthesis often requires months or even years to conclude \citep{tsertsvadze2015, okoli2015}. This protracted timeline inevitably compromises the completeness, timeliness, and reproducibility of the findings. This challenge is particularly acute in quantitative finance, where the research landscape is fragmented across diverse methodological traditions, ranging from classical stochastic calculus to modern machine learning. In such an environment, a single review must synthesize thousands of papers across dense, overlapping conceptual categories, a task that has effectively outpaced human cognitive capacity.

In response to this scalability crisis, bibliometric and network-based methods, such as citation analysis \citep{shi2021} and keyword co-occurrence mapping \citep{lages2023}, have been adopted to provide quantitative field overviews. However, a fundamental gap remains: these methods rely on metadata as a proxy for content, rather than analyzing the research's actual semantic substance. Consequently, they fail to distinguish between papers that share identical keywords but offer fundamentally different scholarly contributions. For instance, a theoretical paper deriving a novel stochastic volatility model and an empirical study testing the Heston model on index options may both utilize keywords like ``option pricing'' and ``stochastic volatility''. Despite this overlap, they belong to distinct research paradigms: one focuses on analytical development while the other on empirical validation. Because metadata-driven clustering is blind to these nuanced methodological distinctions, it remains a ``surface-level'' tool that cannot achieve the deep, content-level understanding required for meaningful academic classification.

While advances in Natural Language Processing (NLP) and Large Language Models (LLMs) offer a potential solution for content-level analysis, significant gaps remain for financial research. Unsupervised topic modeling frameworks such as LDA \citep{blei2003} and BERTopic \citep{grootendorst2022} struggle to produce a structured classification hierarchy due to the dense thematic interconnectedness of financial scholarship. In domains like derivative pricing, a single study often integrates stochastic modeling, numerical computation, and empirical calibration, causing unsupervised methods to collapse distinct methodologies into undifferentiated clusters. Furthermore, these data-driven approaches are inherently sensitive to publication volume. Topics that are numerically underrepresented but conceptually pivotal, such as behavioral pricing paradigms or quantum computing approaches, are frequently absorbed into dominant clusters or overlooked entirely. Our preliminary experiments (Section~\ref{subsec:design_rationale}) confirm that purely data-driven methods fail to align with expert-defined taxonomies, which necessitates an expert-supervised framework capable of capturing the granular distinctions that define high-level financial research.

Beyond these algorithmic limitations, a functional gap exists in the current evaluation landscape. Existing AI-assisted SLR tools \citep{vandeSchoot2021} focus primarily on screening, while prominent FinLLM benchmarks including FinanceBench \citep{islam2023} and the Open Financial LLM Leaderboard \citep{lin2025} are uniformly oriented toward practical tasks like sentiment analysis. Consequently, these models remain decoupled from the nuanced demands of scholarly categorization, which requires the ability to distinguish between closely related yet conceptually distinct research contributions. 

To address these gaps, we propose LR-Robot, a supervised–augmented framework that combines expert-designed classification taxonomies with LLM-based execution and human-in-the-loop evaluation. The core design principle is a clear division of labor: domain experts define \textit{what to classify} by constructing multi-dimensional taxonomies grounded in the research literature, LLMs execute \textit{how to classify} by processing abstracts at scale under expert-designed prompt constraints, and human evaluation determines \textit{how well} the classification performs through systematic accuracy and consistency assessment. This approach is motivated by our finding that unsupervised methods cannot produce meaningful category structures in domains with high terminological overlap (detailed in Section~\ref{subsec:design_rationale}), and by the observation that simple linguistic constraints in prompts can substantially improve LLM classification performance in specialized domains without requiring model fine-tuning or domain adaptation. The framework further integrates retrieval-augmented generation (RAG) to maintain an up-to-date knowledge base that supports downstream analyses such as citation network construction and topic evolution tracking \citep{meloni2023, salatino2022}. We demonstrate and evaluate LR-Robot on a corpus of 12,666 option pricing articles from the Scopus database, designing a four-dimensional classification scheme and systematically comparing five mainstream LLMs across dimensions of varying complexity.

The main contributions of this work are fourfold. First, we provide the first systematic evaluation of LLM capabilities on financial academic literature classification tasks, comparing five models across four dimensions of varying complexity and filling a gap in existing FinLLM benchmarks, which focus exclusively on practical financial tasks and do not address academic literature understanding. Second, we characterize the relationship between task complexity and LLM reliability in domain-specific classification, and demonstrate that expert-designed prompt constraints yield disproportionately large gains without model fine-tuning, offering practical guidance for deploying LLMs in specialized domains. Third, we propose and validate a content-level semantic classification methodology that combines expert-designed taxonomies with LLM execution, showing how this approach overcomes the limitations of unsupervised topic modeling in domains with high terminological overlap. Fourth, we apply the framework to produce the largest multi-dimensional mapping of the option pricing literature to date, and demonstrate that temporal co-occurrence analyses and label-enhanced citation networks can reveal structural research trends that are invisible to traditional bibliometric methods.

The remainder of this paper is organized as follows. Section~\ref{sec.framework} presents the LR-Robot framework, including the design rationale for expert-designed taxonomies and the human-in-the-loop evaluation pipeline. Section~\ref{sec.application} describes the option pricing case study, details the four classification dimensions, and reports the systematic LLM evaluation results based on the selected sample. Section~\ref{sec.full_application} applies the framework to the entire option pricing literature, presenting classification distributions, temporal evolution of research themes, and label-enhanced citation network analyses. Section~\ref{sec.conclusion} discusses the implications, limitations, broader applicability of the findings and concludes with a summary and directions for future research.

\section{Framework: LR-Robot}\label{sec.framework}

\subsection{Design Rationale: Why Expert-Designed Taxonomy?}\label{subsec:design_rationale}

A natural starting point for classifying a large body of academic literature is unsupervised topic modeling, which can automatically discover latent thematic structures without manual annotation. Methods such as Latent Dirichlet Allocation (LDA) \citep{blei2003} and BERTopic \citep{grootendorst2022} have been successfully applied in domains where research sub-areas exhibit relatively distinct vocabularies. For instance, \cite{xue2026} demonstrated that BERTopic can produce coherent topic structures in the AI and computer science literature, where sub-fields such as computer vision, natural language processing, and reinforcement learning employ sufficiently differentiated terminology.
 
\begin{figure}[htbp]
    \centering
    
    \begin{subfigure}[b]{0.48\textwidth}
        \centering
        \includegraphics[width=\textwidth]{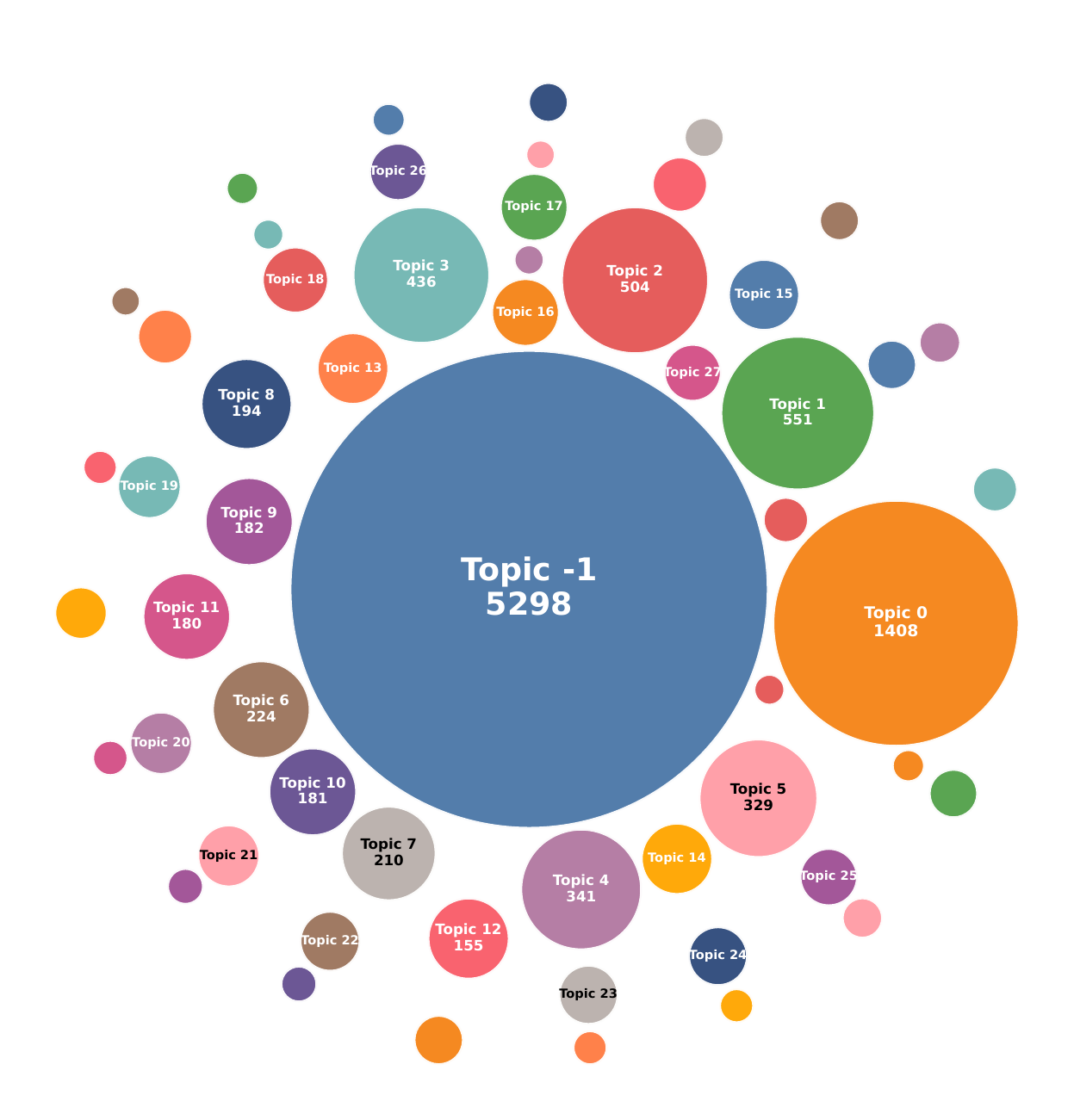}
        \caption{Topic sizes. Bubble area is proportional to document count.}
        
    \end{subfigure}
    \hfill
    \begin{subfigure}[b]{0.48\textwidth}
        \centering
        \includegraphics[width=\textwidth]{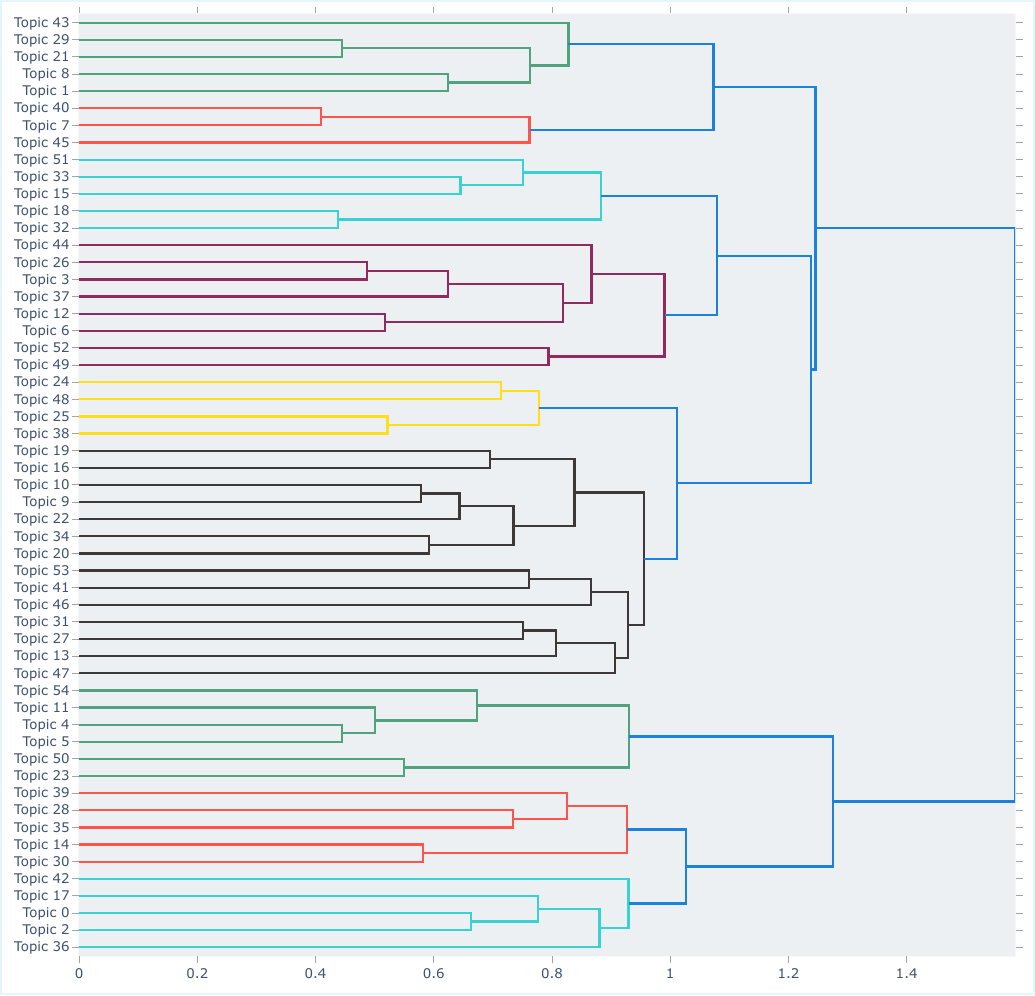}
        \caption{Hierarchical clustering of BERTopic-derived topics}
       
    \end{subfigure}
    
    \caption{Overview of BERTopic results. Left: topic sizes measured by document count. Right: hierarchical clustering illustrating semantic relationships among topics.}
    \label{fig:bertopic_overview}
\end{figure}

Motivated by these advantages, we adopt BERTopic as a baseline and apply it to the option pricing literature following the standard modeling pipeline, which consists of document embedding, dimensionality reduction via Uniform Manifold Approximation and Projection (UMAP), density-based clustering using Hierarchical Density-Based Spatial Clustering of Applications with Noise (HDBSCAN), and topic representation using class-based term frequency--inverse document frequency (c-TF-IDF), with hyperparameter tuning applied to key components. Comprehensive results, including topic outputs, hyperparameter settings, and evaluation metrics, are provided in Appendix~\ref{appendixbertopic}. 

However, our preliminary experiments with BERTopic on the option pricing corpus reveal several fundamental limitations. As shown in Fig.~\ref{fig:bertopic_overview}(a), of the 12,666 abstracts processed, BERTopic identified 54 topics\footnote{The topic indices, keyword representations, and corresponding descriptions are reported in Table~\ref{table:bertopic}.} but assigned nearly half of all papers (5,298) to the outlier category (-1), indicating that the model was unable to place a substantial portion of the literature into any coherent thematic cluster. To examine the relationships among topics, we perform hierarchical clustering on the topic embeddings obtained from BERTopic using cosine distance and an agglomerative clustering procedure, as shown in Fig.~\ref{fig:bertopic_overview}(b). 

The resulting hierarchical structure  exhibits two characteristic failure modes: fragmentation and conflation. Monte Carlo simulation, a single core numerical methodology, is fragmented across three separate topics (Topics~29, 21, and~40) merely because the papers apply it to different option types (American options, general finance, and barrier options respectively), while barrier option pricing is similarly scattered across Topics~40, 7, and~45. At the same time, fundamentally different computational paradigms are conflated into the same branch: Topics~43 (Quantum Algorithms), 8 (GPU-Accelerated Algorithms), 1 (Neural Networks), and 29 (Monte Carlo) are clustered together despite representing distinct methodological approaches. These patterns arise because BERTopic clusters on lexical co-occurrence rather than methodological substance, grouping papers that share similar terminology regardless of their actual research contribution. This mismatch is not a deficiency of the algorithm itself, but a structural incompatibility between unsupervised topic modeling and the characteristics of this domain, where distinctions between research areas are conceptual rather than lexical.

This analysis motivates the core design choice of LR-Robot. Instead of relying on data-driven topic discovery, we adopt an \textit{expert-designed taxonomy}, in which domain experts explicitly define the classification schema, while large language models (LLMs) are used to perform scalable classification. In this framework, the primary role of the expert is not to manually label instances, but to delineate the conceptual boundaries that define meaningful categories—boundaries that cannot be reliably inferred from the data alone.

\subsection{Framework Overview}\label{subsec:framework_overview}
 
LR-Robot is developed to address the challenges of financial academic literature analysis. While designed and validated in the financial domain, the underlying methodology can be adapted to other research fields facing similar challenges of high terminological overlap and complex multi-dimensional classification.
 
\begin{figure}[h]
\centering
\includegraphics[width=1\textwidth]{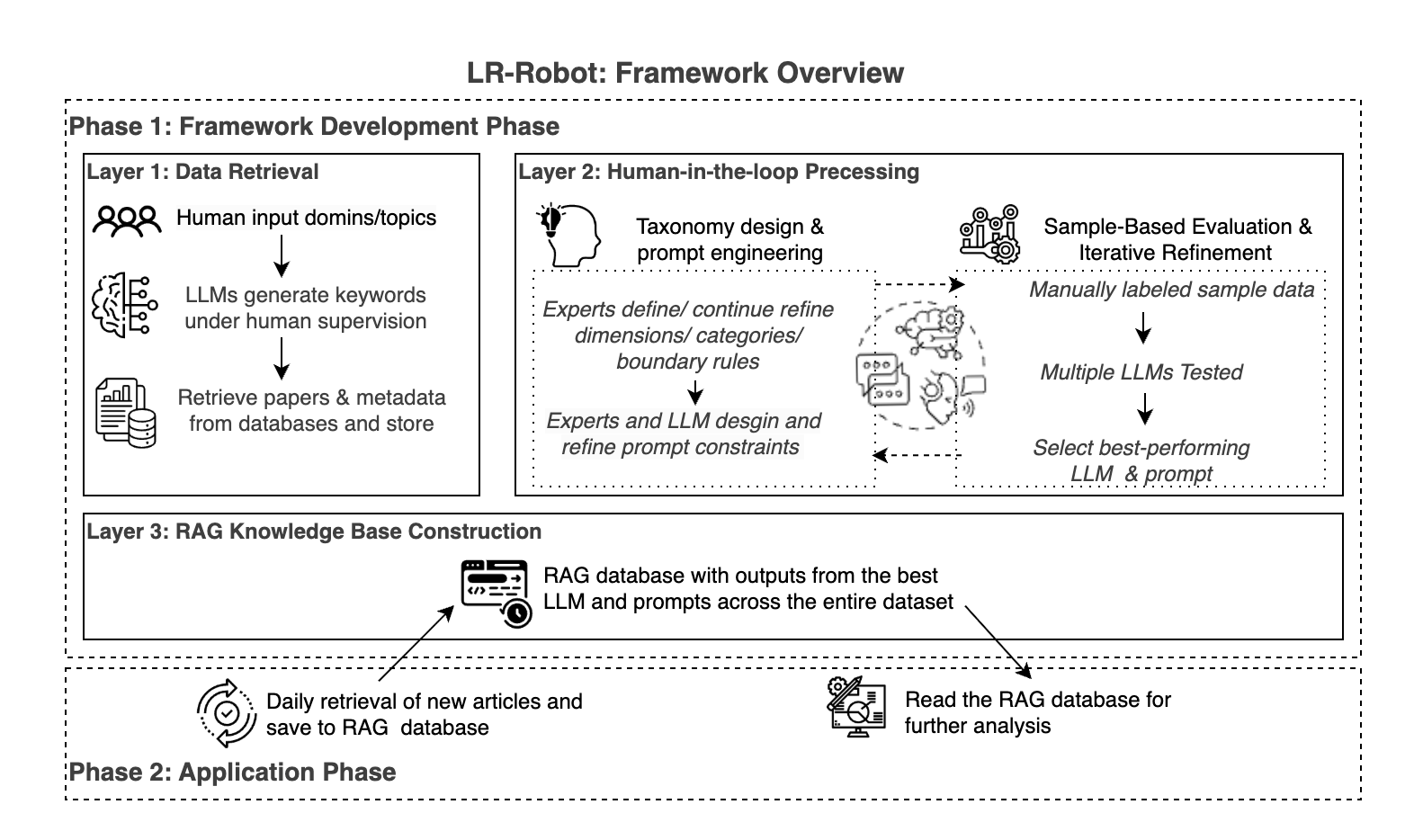}
\caption{Overview of the LR-Robot framework. The three-layer development process (Data Retrieval, Human-in-the-Loop Processing, and RAG Knowledge Base Construction) is shown in the upper portion, with the iterative evaluation loop within Layer~2 indicated by dashed arrows. The lower portion illustrates the application phase, in which new articles are continuously ingested and classified, and users retrieve information from the  RAG knowledge base.}\label{fig:framework}
\end{figure}

The framework comprises three architectural layers (Fig.~\ref{fig:framework}): Data Retrieval, Human-in-the-Loop Processing, and RAG Knowledge Base Construction. Together, these layers define the \textit{Framework Development phase}, a structured process in which domain experts collaborate with LLMs to build a classified, queryable knowledge base for a target research field. 

Once development is complete, the system supports an \textit{Application phase}, which supports the continuous ingestion of newly published articles, information retrieval, and downstream analyses such as citation network construction and topic evolution analysis. The primary contribution of the framework lies in the design of the development process itself, while the application phase illustrates how the resulting knowledge base can be used. A concrete demonstration is provided in Section~\ref{sec.full_application}.

\subsection{Framework Development Phase}\label{subsec:construction}
 
\subsubsection{Layer 1: Data Retrieval}

In the first layer, domain experts collaborate with an LLM to formulate search queries that specify relevant topics, keywords, and inclusion criteria. These queries are then executed across major bibliographic databases (e.g., Scopus, Web of Science) to retrieve relevant records. Depending on data availability and access conditions, the retrieved content may consist of either full-text articles or structured metadata, including titles, abstracts, author information, publication years, and reference lists. All retrieved records are stored in a structured database, which serves as the input corpus for subsequent classification and analysis. The choice of text granularity has implications for downstream classification. While abstracts are often sufficient for capturing high-level attributes, access to full-text content can provide additional detail that may improve the identification of more nuanced methodological distinctions.
 
\subsubsection{Layer 2: Human-in-the-Loop Processing}
 
The second layer is the methodological core of the framework. It encompasses taxonomy design, prompt engineering, and model selection, all embedded within an iterative cycle of expert-guided design, sample-based evaluation, and systematic refinement. Importantly, evaluation is not treated as a discrete, terminal step, but as an integral component of the entire process. At each iteration, domain experts assess LLM outputs on a representative, manually annotated sample and use the results to refine the taxonomy, adjust prompt specifications, and compare candidate models. 
 
\bmhead{Taxonomy Design and Prompt Engineering} Domain experts define classification dimensions and their associated categories, grounded in existing literature and disciplinary conventions. In the option pricing domain, for instance, we define four dimensions (pricing model relevance, underlying asset type, option type, and model type) drawing on established survey frameworks \citep{shrma2025, ruf2020neuralnetworksoptionpricing, broadie2004}. Experts then design prompt constraints that encode domain-specific rules to address boundary cases where LLMs are likely to misclassify (see Appendix~\ref{appendix1} for complete prompt designs). This approach encodes expert knowledge without requiring model fine-tuning. 
 
\bmhead{Sample-Based Evaluation and Iterative Refinement} Because LLM classification performance varies across domains, taxonomies, and prompt configurations, systematic evaluation on a manually labeled representative sample is necessary to validate reliability before full-dataset application. The evaluation results inform taxonomy adjustment, prompt constraint refinement, and model selection simultaneously. This iterative loop between design and evaluation continues until classification quality meets predefined standards. Section~\ref{sec.application} details the specific evaluation metrics and procedures applied in our option pricing case study.
 
\subsubsection{Layer 3: RAG Knowledge Base Construction}
 Once the best-performing model and prompt configuration have been identified, the model generates classification outputs across the entire dataset. These outputs, together with original bibliometric metadata, abstracts, and reference lists, are stored in a structured retrieval-augmented generation (RAG) knowledge base \citep{meloni2023, salatino2022}. This design provides factual grounding by enabling retrieval of classified records rather than reliance on parametric memory, supports structured downstream analysis through multi-dimensional labels attached to each article (enabling label-enhanced citation networks, co-occurrence mapping, and temporal evolution tracking), and ensures transparency and reproducibility by persistently storing all outputs for audit and re-evaluation.

\subsection{Application Phase}\label{subsec:application}
 
Once the framework development is complete, the resulting RAG knowledge base serves as a living infrastructure for ongoing research support. Newly published articles matching the predefined search criteria are retrieved on a regular basis, automatically classified using the validated model and prompt configuration, and appended to the knowledge base without further expert intervention. As the classified corpus grows, it supports a range of downstream uses, from targeted literature retrieval by classification dimension to structural analyses such as label-enhanced citation networks and temporal co-occurrence tracking. Section~\ref{sec.full_application} demonstrates these capabilities through an application to the option pricing literature.

\section{ Framework Development: An Option Pricing Case Study}\label{sec.application}
We apply the three-layer development process of LR-Robot (Section~\ref{sec.framework}) to the literature on option pricing models. Option pricing provides a particularly suitable testbed: the field has accumulated a large volume of publications over decades, its core concepts (e.g., model derivation, numerical solution, and calibration) exhibit substantial terminological overlap that challenges automated classification, and extensive domain expertise is available to validate AI-generated outputs. Our review focuses specifically on papers that develop or compare pricing and volatility models, rather than the broader option pricing literature that includes empirical studies, risk management, and market microstructure. Through systematic evaluation of multiple LLMs across classification tasks of varying complexity, this section establishes, to the best of our knowledge, the first reference benchmarks for LLM accuracy and consistency in financial academic literature classification, providing empirical guidance on where AI can be reliably deployed and where human oversight remains essential.

\subsection{Data Overview}

We collect bibliometric data from the Scopus database, encompassing all English-language research articles, conference papers, and review articles related to option pricing published up to March 5, 2026.\footnote{The data are retrieved using the following query: TITLE-ABS-KEY ("pric* of option*" OR "option pric*" OR "implied volatility" OR "European option*" OR "American option*" OR "stock option*" OR "crypto* option*" OR "interest option*" OR "interest rate option*" OR "exotic option*" OR "basket option*" OR "barrier option*" OR "binary option*" OR "Bermuda option*" OR "compound option*") AND (LIMIT-TO (LANGUAGE, "English")) AND (LIMIT-TO (DOCTYPE, "ar") OR LIMIT-TO (DOCTYPE, "cp") OR LIMIT-TO (DOCTYPE, "re") OR LIMIT-TO (DOCTYPE, "cr")).} The search query combines general terms (e.g., ``option pricing'', ``implied volatility'') with specific contract types (e.g., ``barrier option'', ``Bermuda option'') to ensure broad coverage of both mainstream and niche research areas. The retrieved metadata include article titles, author names, publication years, source titles, abstracts, and reference lists.

The initial search yields 16,174 records. After removing entries with missing or incomplete metadata (primarily absent abstracts, which are required for LLM-based classification), 12,666 valid articles are retained.  These articles span publications from the early 1970s to 2026, reflecting  over five decades of active research. As shown in Fig.~\ref{fig:dim_temporal}(a), annual output has grown from fewer than 50 papers in the early 1990s to over 700 by 2024--2025, consistent with the broader expansion of quantitative finance research 
\citep{donthu2021}.

\subsection{Multi-Dimensional Classification and Evaluation}\label{subsec:classification_evaluation}

We define four classification dimensions for analyzing the option pricing literature, grounded in established survey works in the field \citep{shrma2025,ruf2020neuralnetworksoptionpricing,broadie2004}: whether a paper develops or compares pricing models, what underlying assets it considers, what option types it addresses, and what modeling approaches it employs. These dimensions capture complementary aspects of each paper's  contribution and they also  span a natural gradient of classification difficulty, from a binary decision with definable concept boundaries to multi-label assignments over categories with substantial semantic overlap.

To evaluate LLM performance on these tasks, we randomly select 1,000 papers from the full dataset and manually label them for Dims~1 to 4. From these, 417 papers identified as addressing pricing or volatility model development are carried forward for manual labeling across Dims~2 through 4. All labels are assigned by domain experts. Each LLM is run three times on the same sample, and reported values are averages across runs.

\subsubsection{Dim 1: Option Pricing Models}\label{subsec:dim1}

The first dimension asks a binary question: does the paper develop or compare a pricing or volatility model? Authors typically state their contribution in the abstract, making this question answerable from abstract-level information. While seemingly straightforward, this distinction requires domain knowledge to operationalize. Many papers in the option pricing literature touch on pricing models without making model development their central contribution, for instance, empirical studies that apply the Black-Scholes model to test market efficiency, or risk management analyses that use pricing outputs as inputs. To address this ambiguity, we design expert constraints that explicitly instruct the LLM to classify such papers as negative. These constraints cover common boundary cases including empirical applications, market microstructure, credit risk modeling, hedging strategies, real options, and energy or weather derivatives. The complete list is provided in Appendix~\ref{subsection: appendix_pricing_model_classification}.

We evaluate eleven LLMs across three families (GPT-4o/5/5.2, Gemini Flash 2.0/2.5/3.0, DeepSeek-8B/70B/V3/R1/V3.2), each run three times, reporting Average Accuracy, Average F1 Score, and Self-consistency.\footnote{For the binary classification in Dim~1, Accuracy is the proportion of correctly classified papers. The F1 Score is the harmonic mean of Precision and Recall: $\text{F1} = 2 \times \text{Precision} \times \text{Recall} \,/\, (\text{Precision} + \text{Recall})$, where Precision $= TP/(TP+FP)$ and Recall $= TP/(TP+FN)$, with $TP$, $FP$, and $FN$ denoting true positives, false positives, and false negatives respectively. Self-consistency is the proportion of papers assigned to the same class in all three runs. All reported values are averages across three runs.} All models are evaluated with and without expert constraints. Table~\ref{tab:Q1_performance} presents the results.

\begin{table}[h]
    \centering
    \caption{Dim~1 Performance with and without Expert Constraints.}
    \label{tab:Q1_performance}
    \begin{tabular}{l ccc ccc}
        \toprule
        & \multicolumn{3}{c}{\textbf{Without Constraints}} & \multicolumn{3}{c}{\textbf{With Constraints}} \\
        \cmidrule(lr){2-4} \cmidrule(lr){5-7}
        \textbf{Model} & \textbf{Accuracy} & \textbf{F1} & \textbf{Self-cons.} & \textbf{Accuracy} & \textbf{F1} & \textbf{Self-cons.} \\
        \midrule
        GPT-4o & 0.7110 & 0.6471 & 0.782 & 0.8273 & 0.7906 & 0.914  \\
        GPT-5 &  0.6537 & 0.6379 & 0.939 & 0.8323 & 0.7905 & 0.951  \\
        GPT-5.2  & 0.6847 & 0.6545 & 0.901 &0.8290 & 0.7942& 0.968  \\
        Gemini Flash 2.0  & 0.7281 & 0.7419 & 0.905 & 0.8327 & 0.8152 & 0.947\\
        Gemini Flash 2.5 &  0.6317 & 0.6829 & 0.899 &0.7857 &0.7812 & 0.871\\
        Gemini Flash 3.0 &  0.6407 & 0.6944 & 0.959 &0.8360 & 0.8186 &0.933 \\
        DeepSeek-8B & 0.6750 & 0.6672 & 0.660 & 0.7187 & 0.7090 & 0.672  \\
        DeepSeek-70B & 0.7310 & 0.7102 & 0.684 &0.7700 & 0.6793 & 0.741 \\
        DeepSeek V3 & 0.6470 & 0.3418 & 0.899  & 0.7857 & 0.7010 & 0.978 \\
        DeepSeek R1 & 0.6837 & 0.6484 & 0.880  & 0.8403 & 0.8175 & 0.888 \\
        DeepSeek V3.2 &0.6603 & 0.5150 & 0.966 &0.7813 & 0.6867 & 0.961 \\
        \bottomrule
    \end{tabular}
\end{table}

The results in Table~\ref{tab:Q1_performance} demonstrate that LLMs can classify option pricing papers at the abstract level with considerable accuracy and stability. With expert constraints, the best-performing models achieve F1 scores above 0.81 and self-consistency above 0.94, indicating that authors' abstracts carry sufficient signal for distinguishing model-development papers from those that merely apply existing models. Expert constraints play a critical role in reaching this level of performance: all eleven models improve when constraints are introduced, with gains ranging from moderate (Gemini Flash 2.0, F1 from 0.7419 to 0.8152) to dramatic (DeepSeek V3, F1 from 0.3418 to 0.7010). Without explicit rules that delineate concept boundaries, even capable models frequently misclassify papers that touch on pricing models without making model development their primary contribution. Across model families, we also observe that newer generations tend to outperform their predecessors in both accuracy and consistency, which informs our decision to focus subsequent evaluations on the latest available models. Among all candidates, Gemini Flash 3.0 achieves the strongest overall balance of accuracy, F1, and self-consistency, and is therefore selected as the primary model for full-dataset classification in this dimension.

\subsubsection{Dim 2: Underlying Asset Types}\label{subsec:dim2}

Since our review focuses on pricing and volatility models, Dims~2 through 4 are applied only to the 417 papers positively classified in Dim~1. The second dimension classifies papers by the underlying assets they study: Stocks, Indices, Commodities, Currencies, Interest Rates, Cryptocurrencies, or Not Specified (for papers developing general frameworks without referencing specific assets). Multi-label assignment is permitted, as some studies examine multiple asset types. These categories have clear semantic boundaries, with asset names typically appearing explicitly in abstracts, making this dimension well suited for abstract-level classification. The detailed prompt design is provided in Appendix~\ref{subsection: appendix_underlying}. Based on the Dim~1 evaluation, which showed that newer-generation models generally outperformed their predecessors, and given that several earlier models had been retired or were no longer accessible via API, we focus the remaining evaluations on five models: GPT-5, GPT-5.2, Gemini Flash 2.5, Gemini Flash 3.0, and DeepSeek V3.2 and each model is run three times. 

Since Dims~2 through 4 involve multi-label classification, the evaluation metrics differ from those used in Dim~1. We measure classification quality against human labels using four metrics: Mean Jaccard Similarity, Lenient Accuracy, Micro-averaged F1, and Sample-averaged F1.\footnote{Mean Jaccard Similarity: $\frac{1}{N}\sum_{i=1}^{N} |y_i \cap \hat{y}_i| / |y_i \cup \hat{y}_i|$, where $y_i$ and $\hat{y}_i$ are the true and predicted label sets for sample $i$. Lenient Accuracy: the proportion of samples for which the model correctly identifies at least one human-labeled category. Sample-F1 computes F1 per sample and then averages: $\frac{1}{N}\sum_{i=1}^{N} 2|y_i \cap \hat{y}_i| / (|y_i| + |\hat{y}_i|)$. Micro-F1 aggregates true positives, false positives, and false negatives globally across all classes: $2TP/(2TP + FP + FN)$.  All accuracy values are reported as mean $\pm$ standard deviation across three runs.} We assess self-consistency using two measures: Full Agreement Rate, the proportion of samples receiving identical label sets across all three runs, and Pairwise Jaccard, the average Jaccard similarity between label sets from each pair of runs, averaged across samples.\footnote{Pairwise Jaccard: for each sample, compute $|L_a \cap L_b| / |L_a \cup L_b|$ for all run pairs $(a,b)$, average across pairs, then average across samples.} We also include a text-mapping baseline that assigns labels based on keyword matching in abstracts.

Table~\ref{tab:Q2_model_comparison} presents the results across models and the text-mapping baseline. All five LLM models achieve Sample F1 above 0.82 on this dimension,  confirming that underlying asset types can be reliably classified from abstracts alone when category boundaries are semantically clear. Notably, all LLMs outperform the text-mapping baseline (Sample F1 0.8273), indicating that even for relatively explicit categories, semantic understanding provides added value beyond keyword matching. Among the models, GPT-5 and DeepSeek V3.2 achieve the highest accuracy (Sample F1 0.8850 and 0.8845), while Gemini Flash 3.0 achieves the highest self-consistency (Pairwise Jaccard 0.9859). Gemini Flash 2.5 is notably weaker in both accuracy (0.8222) and consistency (0.9307).

\begin{table}[htbp]
    \centering
    \caption{Dim~2: Accuracy and Self-Consistency Across Models.}
    \label{tab:Q2_model_comparison}
    \begin{tabular}{lcccccc}
        \toprule
        & \multicolumn{4}{c}{\textbf{Accuracy}} & \multicolumn{2}{c}{\textbf{Self-consistency}} \\
        \cmidrule(lr){2-5} \cmidrule(lr){6-7}
        \textbf{Model} & \textbf{Jaccard} & \textbf{Len. Acc.} 
        & \textbf{Samp. F1} & \textbf{Micro F1} & \textbf{Full Ag.} 
        & \textbf{Pw. Jac.} \\
        \midrule
        GPT-5 & 0.8829 & 0.8897 & 0.8850 & 0.8824 
        & 0.9760 & 0.9856 \\
        GPT-5.2 & 0.8666 & 0.8793 & 0.8706 & 0.8663 
        & 0.9592 & 0.9771 \\
        Gemini Flash 2.5 & 0.8136 & 0.8401 & 0.8222 & 0.8135 
        & 0.8801 & 0.9307 \\
        Gemini Flash 3.0 & 0.8670 & 0.8809 & 0.8714 & 0.8649 
        & 0.9712 & 0.9859 \\
        DeepSeek V3.2 & 0.8809 & 0.8921 & 0.8845 & 0.8786 
        & 0.9568 & 0.9734 \\
        \midrule
        Text Mapping & 0.8173 & 0.8489 & 0.8273 & 0.8152 
        & --- & --- \\
        \bottomrule
    \end{tabular}
\end{table}

\subsubsection{Dim 3: Option Types}\label{subsec:dim3}

The third dimension classifies the same 417 papers by option type: European, American, Exotic, or Not Specified. Papers that do not explicitly mention exotic options but discuss related instruments such as Asian, Barrier, or Basket options are classified as Exotic. As with Dim~2, multi-label assignment is permitted, and we evaluate the same five models using the accuracy and self-consistency metrics defined in Section~\ref{subsec:dim2}. Option type names are among the most explicit textual cues in financial abstracts, making this the most straightforward multi-label task in our framework and we also include a simple text-mapping baseline for comparison.

Table~\ref{tab:Q3_model_comparison} presents the results across models. Since option type names appear as near-exact keywords in abstracts, we include a text-mapping baseline for comparison. Text mapping achieves the highest accuracy on this dimension (Sample F1 0.9251), surpassing all LLMs, which reflects the keyword-like nature of option type terminology. However, text mapping is susceptible to false positives when a keyword appears in a context unrelated to the paper's focus, for example, a paper mentioning ``European option'' only as a benchmark rather than as its subject. Apart from Gemini Flash 2.5, which is again the weakest (Sample F1 0.8225, Pairwise Jaccard 0.9079), the remaining four models achieve comparable accuracy and consistency, with differences within two percentage points.

\begin{table}[htbp]
    \centering
    \caption{Dim~3: Accuracy and Self-Consistency Across Models.}
    \label{tab:Q3_model_comparison}
    \begin{tabular}{lcccccc}
        \toprule
        & \multicolumn{4}{c}{\textbf{Accuracy}} & \multicolumn{2}{c}{\textbf{Self-consistency}} \\
        \cmidrule(lr){2-5} \cmidrule(lr){6-7}
        \textbf{Model} & \textbf{Jaccard} & \textbf{Len. Acc.} 
        & \textbf{Samp. F1} & \textbf{Micro F1} & \textbf{Full Ag.} 
        & \textbf{Pw. Jac.} \\
        \midrule
        GPT-5 & 0.8812 & 0.9241 & 0.8955 & 0.8933 
        & 0.9424 & 0.9742 \\
        GPT-5.2 & 0.8873 & 0.9265 & 0.9001 & 0.8990 
        & 0.9760 & 0.9864 \\
        Gemini Flash 2.5 & 0.8099 & 0.8489 & 0.8225 & 0.8304 
        & 0.8465 & 0.9079 \\
        Gemini Flash 3.0 & 0.8680 & 0.9137 & 0.8829 & 0.8826 
        & 0.9400 & 0.9670 \\
        DeepSeek V3.2 & 0.8820 & 0.9169 & 0.8933 & 0.8938 
        & 0.9305 & 0.9607 \\
        \midrule
        Text Mapping & 0.9169 & 0.9424 & 0.9251 & 0.9221 
        & --- & --- \\
        \bottomrule
    \end{tabular}
\end{table}

\subsubsection{Dim 4: Option Model Types} \label{Option_Model_Types}
In the fourth dimension, we classify the papers based on the type of option pricing model described in their abstracts. The classification framework was developed through extensive consultation involving AI-assisted analyses and expert domain discussions, resulting in eight primary categories: (1) Analytical Models, (2) Numerical Methods, (3) Multi-Factor and Hybrid Models, (4) Market Imperfections and Frictions, (5) Calibration and Model Estimation, (6) Machine Learning and Data-Driven Approaches, (7) Behavioral and Alternative Paradigms, (8) Emerging and Niche Approaches or Others. Each category encompasses more fine-grained subclasses that capture specific modeling nuances within the broader conceptual framework. The detailed prompt design and classification procedure are documented in Appendix \ref{subsection: appendix_pricing_model_types}. As with Dims~2 and 3, multi-label assignment is permitted, and we evaluate the same five models using the metrics defined in Section~\ref{subsec:dim2}.

In Table~\ref{tab:Q4_model_comparison}, ``Direct'' refers to the model's class-level response when asked to classify into the eight major categories directly, while ``Sub$\rightarrow$Class'' derives class labels by first classifying into the 33 subclasses and then mapping upward. The rationale is that finer-grained prompts may force the model to engage more precisely with methodological details in the abstract before committing to a broad category. The results confirm this: Sub$\rightarrow$Class consistently outperforms Direct across all models and all metrics, with Sample F1 improving by 0.08 to 0.12 and Full Agreement Rate by 0.05 to 0.17. We therefore adopt the finer-grained subclass-level prompt for Dim~4 when classifying the full dataset. 

Regarding model selection, both GPT-5.2 and Gemini Flash 3.0 emerge as strong candidates: GPT-5.2 achieves the highest accuracy (Jaccard 0.5627, Micro F1 0.6688, Full Agreement 0.8034), while Gemini Flash 3.0 achieves the highest Lenient Accuracy (0.8449) and Pairwise Jaccard (0.9389), meaning it rarely misses relevant categories and produces the most stable outputs across runs. DeepSeek V3.2 achieves competitive accuracy but substantially lower consistency (Full Agreement 0.5204).

\begin{table}[htbp]
    \centering
    \caption{Dim~4: Class-Level Accuracy and Self-Consistency 
    Across Models (Direct vs. Subclass-Derived).}
    \label{tab:Q4_model_comparison}
    \begin{tabular}{llcccccc}
        \toprule
        & & \multicolumn{4}{c}{\textbf{Accuracy}} 
        & \multicolumn{2}{c}{\textbf{Self-consistency}} \\
        \cmidrule(lr){3-6} \cmidrule(lr){7-8}
        \textbf{Model} & \textbf{Method}
        & \textbf{Jac.} & \textbf{Len.} 
        & \textbf{Sa.F1} & \textbf{Mi.F1} 
        & \textbf{F.Ag.} & \textbf{Pw.J.} \\
        \midrule
        \multirow{2}{*}{GPT-5} 
        & Direct & 0.4432 & 0.7634 & 0.5340 & 0.5445 
        & 0.7050 & 0.9050 \\
        & Sub$\rightarrow$Class & 0.5364 & 0.8114 & 0.6176 & 0.6371 
        & 0.7146 & 0.9119 \\
        \midrule
        \multirow{2}{*}{GPT-5.2} 
        & Direct & 0.4418 & 0.7890 & 0.5396 & 0.5494 
        & 0.7194 & 0.9098 \\
        & Sub$\rightarrow$Class & 0.5627 & 0.8329 & 0.6436 & 0.6688 
        & 0.8034 & 0.9380 \\
        \midrule
        \multirow{2}{*}{Gemini Flash 2.5} 
        & Direct & 0.4291 & 0.8225 & 0.5364 & 0.5408 
        & 0.4988 & 0.8542 \\
        & Sub$\rightarrow$Class & 0.5349 & 0.8377 & 0.6242 & 0.6462 
        & 0.6715 & 0.8996 \\
        \midrule
        \multirow{2}{*}{Gemini Flash 3.0} 
        & Direct & 0.4183 & 0.8297 & 0.5305 & 0.5381 
        & 0.6043 & 0.8910 \\
        & Sub$\rightarrow$Class & 0.5179 & 0.8449 & 0.6141 & 0.6360 
        & 0.7698 & 0.9389 \\
        \midrule
        \multirow{2}{*}{DeepSeek V3.2} 
        & Direct & 0.4212 & 0.7514 & 0.5139 & 0.5212 
        & 0.3573 & 0.7523 \\
        & Sub$\rightarrow$Class & 0.5575 & 0.8090 & 0.6327 & 0.6561 
        & 0.5204 & 0.8294 \\
        \bottomrule
    \end{tabular}
\end{table}

\subsection{Error Distribution Analysis}

\begin{figure}[htbp]
\centering
\includegraphics[width=\textwidth]{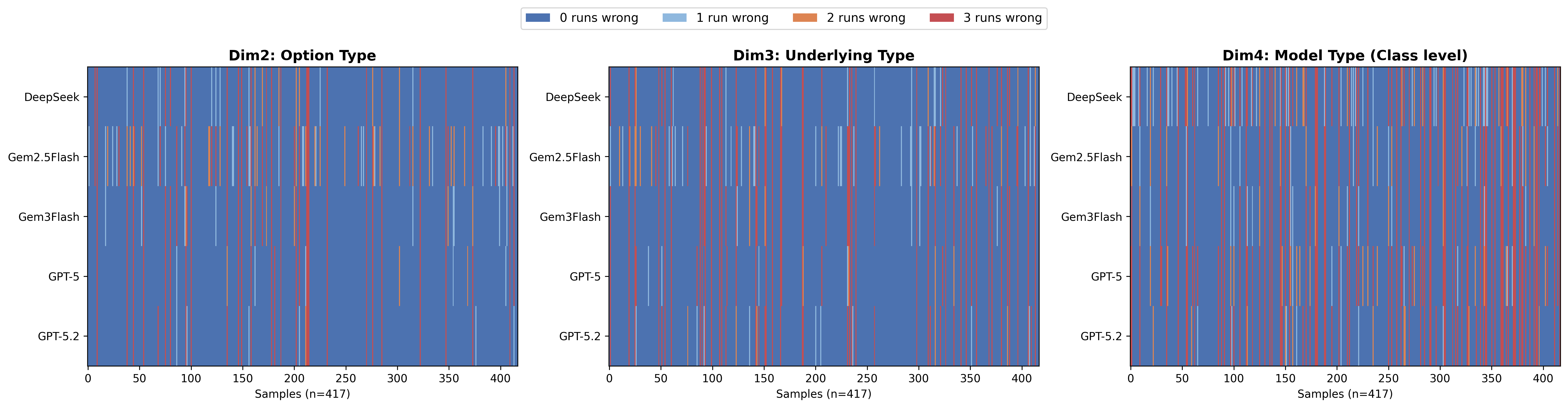}
\caption{Per-sample error patterns across five models and three  dimensions. Top: each cell indicates whether a model classifies a paper correctly in all three runs (dark blue), one run wrong (light blue), two runs wrong (orange), or all three runs wrong (red). Bottom: total error count per paper aggregated across all models.}
\label{fig:error_heatmap}
\end{figure}

To understand whether classification errors are driven by individual model weaknesses or by inherent ambiguity in the papers, we visualize the per-sample, per-model error pattern across all three multi-label dimensions. For each of the 417 papers and each model, we record the number of runs that produce a complete misclassification (no overlap with human labels, i.e., Lenient Accuracy $= 0$ for that sample. Fig.~\ref{fig:error_heatmap} displays the results, with the bottom panels aggregating the total error count per paper across all models.

Across Dims~2 to 4, errors exhibit the same structural pattern: vertical red stripes spanning all five models, indicating that misclassifications concentrate on the same papers regardless of which model is used. This consistency across models suggests that errors are driven by inherent ambiguity in the abstracts, such as insufficient methodological detail or papers that genuinely straddle multiple categories, rather than by idiosyncratic model weaknesses. The difference across dimensions is one of frequency: Dims~2 and 3 show only occasional red stripes, while Dim~4 shows substantially more, consistent with the greater conceptual overlap among model-type categories. 

Beyond aggregate accuracy and self-consistency, the error heatmap also informs model selection. A model that errs primarily on universally hard papers (where all models fail) is preferable to one that additionally produces unique errors on otherwise easy papers. Examining the model-specific columns in Fig.~\ref{fig:error_heatmap}, Gemini Flash 3.0 and GPT-5.2 exhibit the fewest isolated errors outside the universally hard subset, suggesting that their mistakes are largely confined to genuinely ambiguous cases. Gemini Flash 3.0 shows the cleanest pattern for Dims~2 and 4, while GPT-5.2 performs best on Dim~3. Given that Dim~4 is the most demanding task, Gemini Flash 3.0 also performs strongly on Dim~2, and it maintains high accuracy and self-consistency across all dimensions (Tables~\ref{tab:Q2_model_comparison}--\ref{tab:Q4_model_comparison}), we select Gemini Flash 3.0 as the primary model for full-dataset classification.

\section{Application to Option Pricing Literature}\label{sec.full_application}

In this section, we apply the best-performing model Gemini Flash 3.0 and the prompt refined through human-in-the-loop instruction to the full dataset. We perform analyses, including category classifications across four dimensions, the evolution of option pricing literature and citation network analysis, demonstrating the effectiveness and versatility of our framework.

\subsection{Research Landscape and Thematic Evolution}
\subsubsection{Temporal Distribution of Research Themes}\label{subsubsec:temporal_dist}

\begin{figure}[htbp]
\centering
\begin{minipage}{\textwidth}
    \centering
    \includegraphics[width=\textwidth]{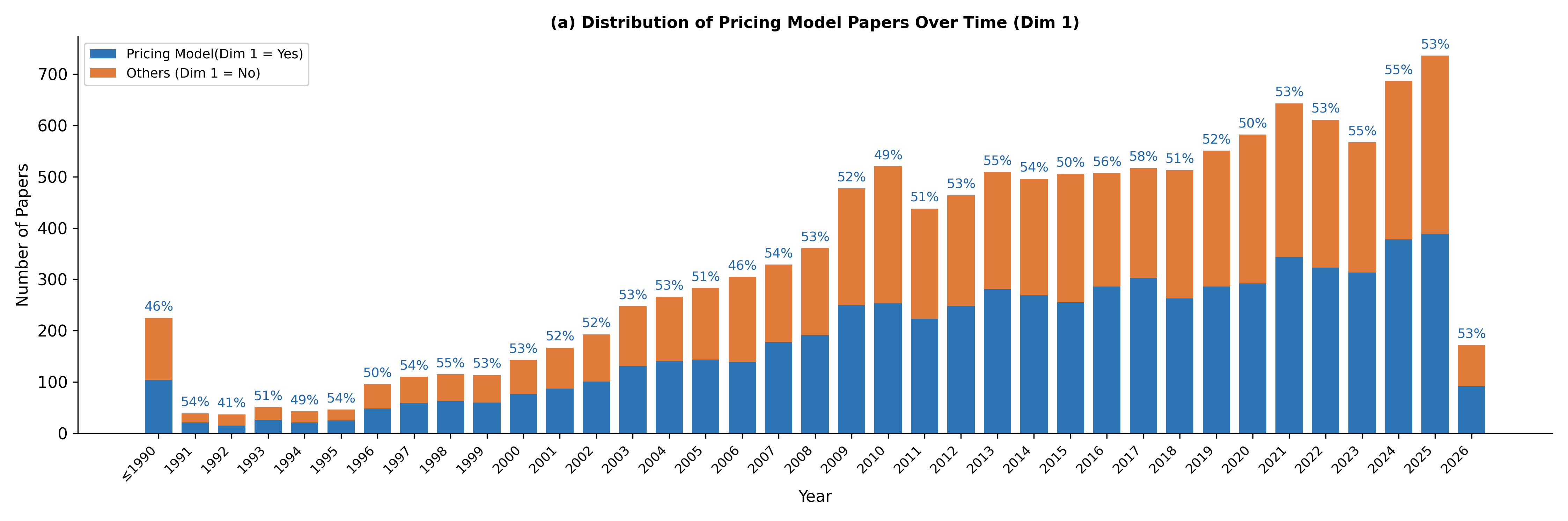}
\end{minipage}
\vspace{0.3cm}
\begin{minipage}{\textwidth}
    \centering
    \includegraphics[width=\textwidth]{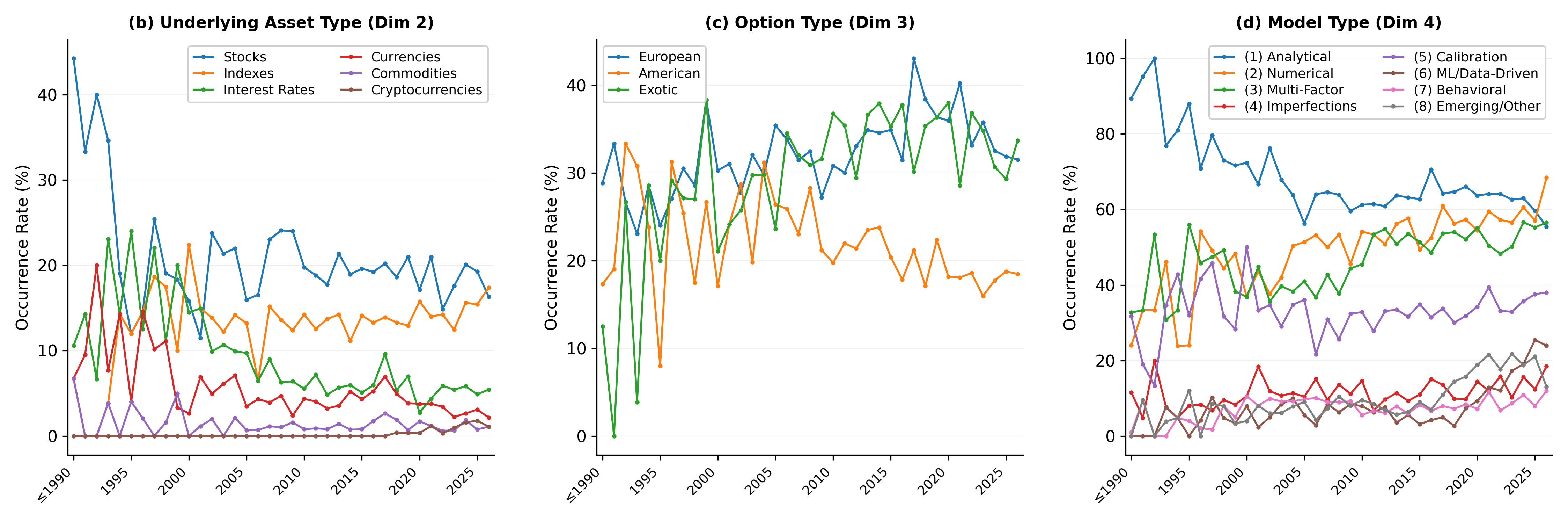}
\end{minipage}
\caption{Temporal distribution of the four classification
dimensions. (a)~Annual publication volume and proportion of
pricing/volatility model papers (Dim~1). (b)--(d)~Occurrence rates
of each label within the 6,766 Dim~1-positive papers for underlying
asset type (Dim~2), option type (Dim~3), and model type (Dim~4),
respectively. Dim~2--4 allow multi-label assignments, so rates can
exceed 100\%. Years prior to 1991 are grouped as ``$\leq$1990''.}
\label{fig:dim_temporal}
\end{figure}

Fig.~\ref{fig:dim_temporal} presents the temporal distribution of all four dimensions. Panel~(a) shows annual publication volume and the Dim~1 modeling share; we identified 6,766 of 12,666 papers (53.42\%) as focusing on pricing or  volatility model development and comparison. Panels~(b)--(d) show occurrence rates for Dim~2--4 within these 6,766 papers. Since Dim~2--4 allow multi-label assignments, rates can exceed 100\%.

\bmhead{Dim~1: Option pricing model}

Despite the rapid growth in publication volume, the share of pricing/volatility model papers has remained stable at approximately 50--55\% throughout the entire period (Fig.~\ref{fig:dim_temporal}a), indicating a persistent equilibrium between theoretical model development and empirical or applied research in option pricing.

\bmhead{Dim~2: Underlying asset types}

Stocks dominated the early literature (over 40\% before 1992) but have declined steadily to roughly 17\% by 2025, reflecting asset-class diversification rather than diminished interest. Indexes maintain a stable 10--15\%, while Interest Rates and Currencies sustain modest presence. Cryptocurrencies appear only after 2018 and remain marginal.

\bmhead{Dim~3: Option types}

European and American options maintain stable occurrence rates of approximately 30\% and 20--30\%, respectively. Exotic options show the clearest trend, rising from under 10\% in the early 1990s to approximately 25\% by 2025, driven by the proliferation of structured products in financial markets.

\bmhead{Dim~4: Model types}

Analytical Models~(1) and Numerical Methods~(2) have dominated throughout at 60--80\%, though both show gradual relative decline as the methodological landscape has diversified. The most significant shift is the rise of ML and Data-Driven Approaches~(6), from near zero before 2005 to approximately 25\% by 2025, coinciding with advances in deep learning and the increasing availability of high-frequency financial data. Multi-Factor Models~(3) have also grown steadily to roughly 35\%, reflecting the need to capture empirically observed features such as stochastic volatility and jumps that single-factor models cannot accommodate. The remaining categories each stay below 15\%.

\subsubsection{Cross-Dimensional Co-occurrence and Its Evolution}\label{subsubsec:cross_dim}

\begin{figure}[htbp]
\centering
\begin{minipage}[t]{0.49\textwidth}
    \centering
    \includegraphics[width=\textwidth]{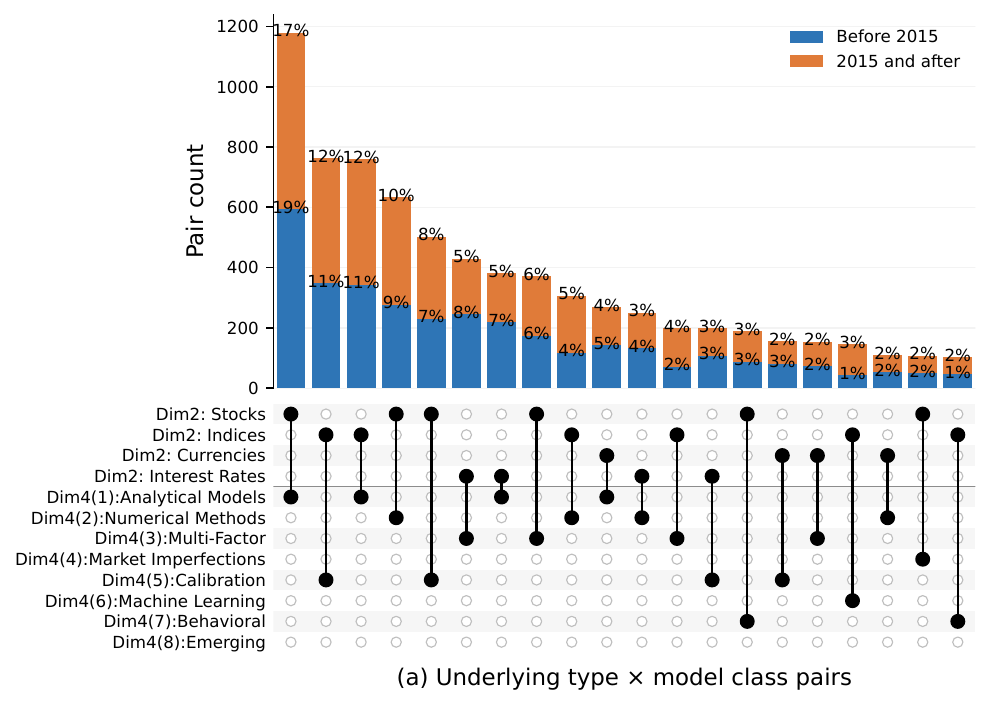}
\end{minipage}
\hfill
\begin{minipage}[t]{0.49\textwidth}
    \centering
    \includegraphics[width=\textwidth]{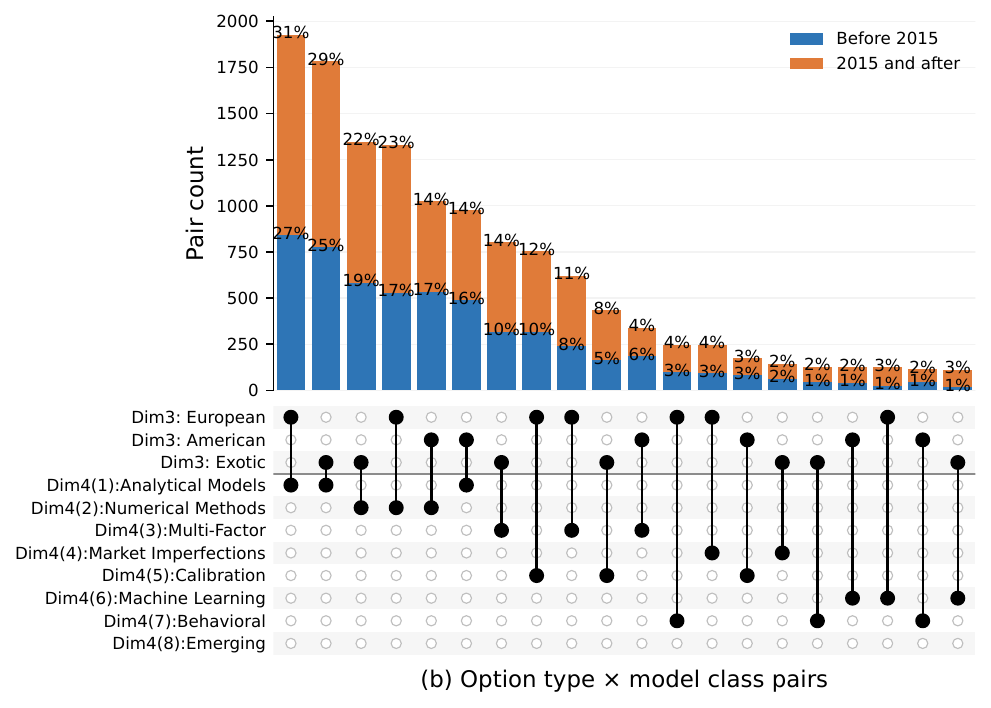}
\end{minipage}
\caption{Top-20 cross-dimensional co-occurrence pairs, split by time period (before 2015 vs.\ 2015 and after, corresponding to the most recent decade of research). Each bar shows the absolute pair count; percentages above each segment indicate the pair's share of total papers in that period. (a)~Underlying type $\times$ model class (Dim~2 $\times$ Dim~4). (b)~Option type $\times$ model class (Dim~3 $\times$ Dim~4).}
\label{fig:cross_dim}
\end{figure}

Fig.~\ref{fig:cross_dim} presents the top-20 co-occurrence pairs for Dim~2$\times$Dim~4 and Dim~3$\times$Dim~4, split by time period (before 2015 vs.\ 2015 and after). The percentage annotations indicate each pair's share of total papers in the corresponding period. Some patterns emerge from the comparison.

\bmhead{Equity underlyings remain stable, interest-rate pairings decline} In Fig.~\ref{fig:cross_dim}(a), the top five positions are occupied by Stocks and Indexes: Stocks$\times$Analytical (18.8\%$\to$16.6\%), Indexes$\times$Calibration (11.0\%$\to$11.8\%), Indexes$\times$Analytical (10.9\%$\to$11.8\%), Stocks$\times$Numerical (8.8\%$\to$10.1\%), and Stocks$\times$Calibration (7.3\%$\to$7.8\%). These proportions remain largely stable across both periods, confirming that equity-linked underlyings studied through classical approaches constitute the enduring core of the field. In contrast, all Interest Rate pairings decline: Interest Rates$\times$Multi-Factor from 7.8\% to 5.2\%, Interest Rates$\times$Analytical from 6.9\% to 4.6\%, and Interest Rates$\times$Numerical from 4.2\% to 3.3\%, consistent with the single-dimension trend observed in Fig.~\ref{fig:dim_temporal}(b).

\bmhead{European and Exotic options drive growth, American options declines}
In Fig.~\ref{fig:cross_dim}(b), the four most frequent combinations all involve European or Exotic options: European$\times$Analytical (26.6\%$\to$30.8\%), Exotic$\times$Analytical (24.6\%$\to$28.6\%), Exotic$\times$Numerical (18.5\%$\to$21.6\%), and European$\times$Numerical (16.7\%$\to$22.8\%). All four show growth in the post-2015 period, with Exotic options now rivalling European options as a primary driver of methodological development. In contrast, American option pairings with the same core methods decline: American$\times$Numerical drops from 16.9\% to 14.0\%, and American$\times$Analytical from 15.5\% to 13.9\%. This suggests that American option pricing has reached a stage of relative methodological maturity, with the expanding complexity of exotic products attracting a growing share of research effort.

\bmhead{ML approaches demonstrate rapid single-dimension growth }
ML-related combinations appear only at the bottom of both Top-20 lists, with shares of 1--3\% in the post-2015 period: European$\times$ML (0.8\%$\to$2.8\%), American$\times$ML (1.3\%$\to$2.4\%), Exotic$\times$ML (0.6\%$\to$2.6\%), and Indexes$\times$ML (1.4\%$\to$2.9\%). Despite ML reaching approximately 25\% as a single-dimension occurrence rate in Fig.~\ref{fig:dim_temporal}(d), its cross-dimensional shares remain low and notably uniform across option types and asset classes. This indicates that current ML research in option pricing is predominantly methodological, developing general-purpose algorithms rather than targeting specific products or assets. The application of ML to exotic option pricing, commodity derivatives, and interest-rate products remains largely unexplored and represents a significant opportunity for future research.

\subsubsection{Cross-Dimensional Co-occurrence and Its Evolution}
\label{subsubsec:cross_dim}

Fig.~\ref{fig:cross_dim} presents the top-20 co-occurrence pairs
for Dim~2$\times$4 and Dim~3$\times$4, split by time period
(before 2015 vs.\ 2015 and after). The percentage annotations
indicate each pair's share of total papers in the corresponding
period. Four patterns emerge from the comparison.

\bmhead{Dominance of Analytical and Numerical pairings}

The leading combinations in both panels are dominated by pairings with Analytical Models and Numerical Methods. In Fig.~\ref{fig:cross_dim}(a), Stocks paired with Analytical and Numerical methods occupy the top positions, followed by similar pairings with Indices and Interest Rates. In Fig.~\ref{fig:cross_dim}(b), the six most frequent pairs are European with Analytical (27\%$\to$31\%), European with Numerical (25\%$\to$29\%), American with Analytical (19\%$\to$22\%),  American with Numerical (17\%$\to$23\%), and Exotic with both Analytical and Numerical (each 10\%$\to$14\%). Notably, the percentage shares of these dominant pairings remain largely stable across the two time periods, confirming that Analytical Models and Numerical Methods have served as the default methodological choices across all underlying asset classes and option types for at least three decades. While these dominant pairings show no sign of displacement, the emergence of new combinations in the lower-ranked positions signals a gradual diversification of the methodological landscape.

\bmhead{Rapid growth of ML in isolation}

Despite reaching approximately 25\% occurrence rate as a single-dimension label (Fig.~\ref{fig:dim_temporal}d), ML-related cross-dimensional combinations remain sparse in the Top~20, with most appearing at only 1--3\% after 2015. This contrast indicates that current ML research in option pricing is predominantly methodological, developing general-purpose pricing algorithms rather than targeting specific asset classes or option types. The application of ML-based approaches to under-explored areas such as exotic option pricing, interest-rate derivatives, and commodity
options represents a significant opportunity for future work.


\subsubsection{Global Citation Network}\label{subsubsec:global}

We constructed citation networks based on the reference information available for 12,560 of the 12,666 papers (99.2\%). We first compute PageRank centrality \citep{page1999} on the global network encompassing all 12,666 papers to identify the most influential articles. To examine whether pricing model research constitutes the intellectual core of the broader option pricing literature, we additionally constructed a closed citation network restricted to the 6,766 Dim~1-positive papers and compared its PageRank \footnote{We apply damping factor 0.85 for computing all PageRank ranking.} with that of the global network. 

\begin{figure}[htbp]
\centering
\begin{minipage}[c]{0.3\textwidth}
    \centering
    \footnotesize
    \setlength{\tabcolsep}{4pt}
    \begin{tabular}{cl}
    \toprule
    \textbf{Glob. / Mod. Rank} & \textbf{Paper} \\
    \midrule
    \multicolumn{2}{l}{\textit{Overlapping papers (in both Top 10)}} \\
    \midrule
    1 / 1   & \cite{black1973}    \\
    2 / 5   & \cite{cox1979}      \\
    3 / 3   & \cite{merton1976}   \\
    4 / 2   & \cite{heston1993}   \\
    5 / 4   & \cite{cox1976}      \\
    6 / 7   & \cite{merton1974}   \\
    7 / 6   & \cite{hull1987}     \\
    8 / 8   & \cite{harrison1979} \\
    \midrule
    \multicolumn{2}{l}{\textit{In Global Top 10 only}} \\
    \midrule
    9 / 11  & \cite{rubinstein1994} \\
    10 / 18 & \cite{geske1979}      \\
    \midrule
    \multicolumn{2}{l}{\textit{In Modeling Top 10 only}} \\
    \midrule
    15 / 9  & \cite{broadie1997}  \\
    14 / 10 & \cite{harrison1981} \\
    \bottomrule
    \end{tabular}
\end{minipage}%
\hfill
\begin{minipage}[c]{0.50\textwidth}
    \centering
    \includegraphics[width=\textwidth]{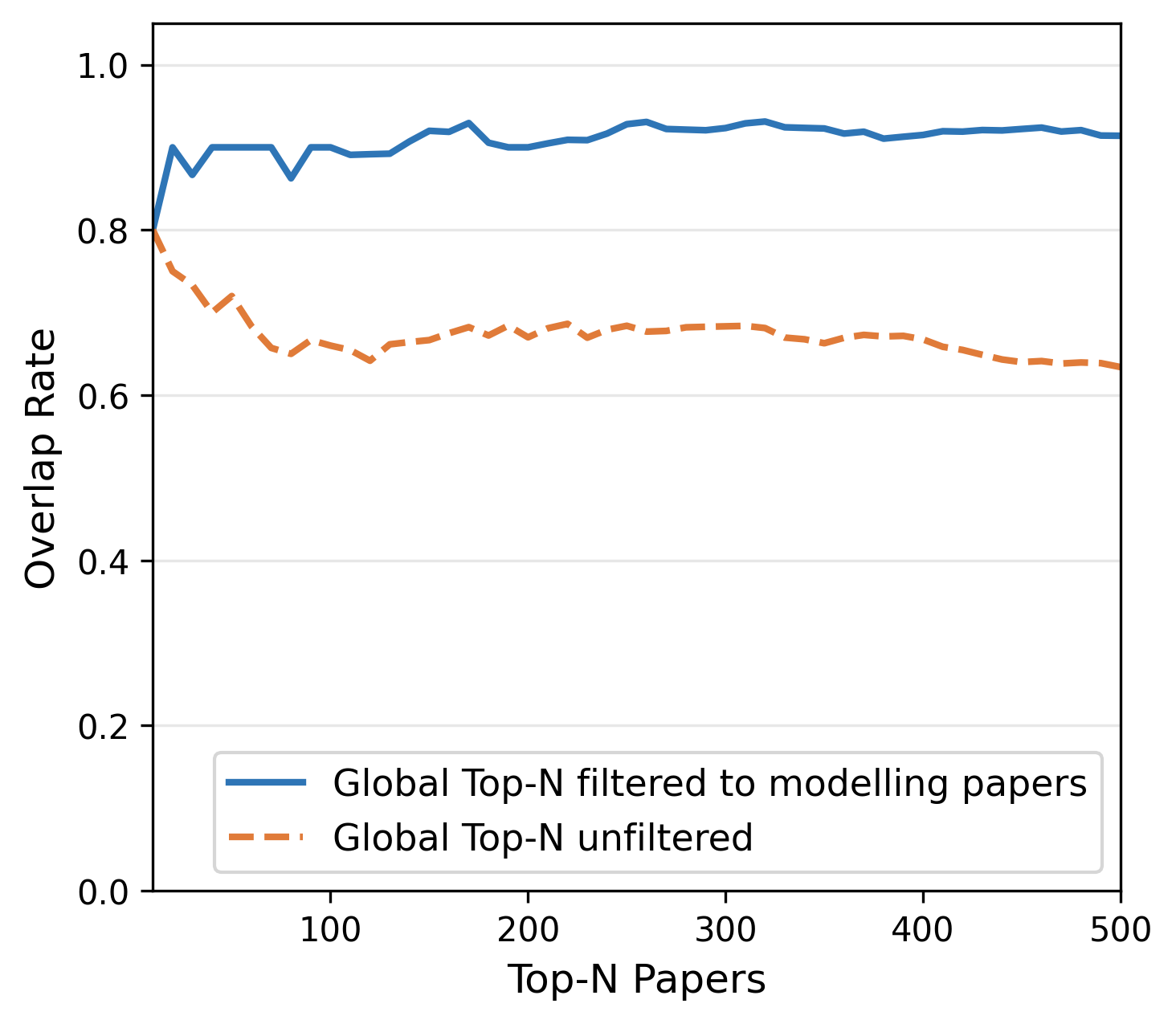}
\end{minipage}

\caption{Comparison of the global (12,666 papers) and modeling (6,766 papers) citation networks. Left: PageRank Top~10 overlap (8/10 coverage). Right: Top-$N$ overlap rate. Solid line restricts the global Top-$N$ to pricing-model papers; dashed line uses the unfiltered global ranking.}
\label{fig:global_vs_modeling}
\end{figure}

As shown in Fig.~\ref{fig:global_vs_modeling} (left), the two Top~10 rankings share 8 out of 10 papers. The overlapping works (\cite{black1973}, \cite{heston1993}, \cite{merton1976}, \cite{cox1979}, \cite{cox1976}, \cite{merton1974}, \cite{hull1987}, and \cite{harrison1979}) are foundational contributions whose influence extends equally across modeling and non-modeling research.

Fig.~\ref{fig:global_vs_modeling} (right) extends this comparison beyond the Top~10 by plotting the overlap rate between the two networks' PageRank Top-$N$ lists as $N$ increases from 10 to 500. A direct comparison of the global Top-$N$ with the modeling Top-$N$ (dashed line) yields an overlap of roughly 65\%. This gap partly arises because the global network ranks all 12,666 papers, so non-modeling works can occupy Top-$N$ positions and reduce the overlap mechanically. To isolate the effect of network structure from that of set composition, we re-rank only the 6,766 pricing-model papers by their global-network PageRank scores and compare the resulting list with the modeling Top-$N$ (solid line). This restricted overlap stabilises above 80\%, confirming that, within the same population of papers, the two networks produce highly consistent importance rankings. The gap between the two curves is itself informative: it indicates that fewer than 20\% of the most influential papers in the global network are non-modeling works, further confirming that pricing model research constitutes the intellectual core of the option pricing field.

These results justify restricting the subsequent label-enhanced analyses to the 6,766-paper modeling network, whose citation structure faithfully reflects the intellectual hierarchy of the entire field while providing the semantic labels necessary for sub-domain analysis. However, the resulting rankings remain uniform across all research directions. In the following subsection, we leverage the multi-dimensional labels produced by LR-Robot to identify works that are distinctively important to each sub-domain.

\subsubsection{{Sub-network Divergence from Global Rankings}}
\label{subsubsec:label_enhanced}

The modeling citation network established in Section~\ref{subsubsec:global} treats the modeling literature as a homogeneous body. We now leverage the multi-dimensional labels produced by LR-Robot to construct category-specific citation sub-networks, restricting source nodes to papers carrying a given thematic label while retaining all 6,766 modeling papers as potential targets. We examine whether these sub-networks reveal citation priorities that differ from the global consensus. All analyses in this subsection are based on the Top-200 papers by PageRank. This threshold balances two considerations: it ensures adequate representation for smaller categories (e.g., Behavioral~(7) contributes only 2 papers to the global Top-200), while remaining selective at approximately the top 3\% of the 6,766-paper network.

Table~\ref{tab:overlap} reports the overlap rate between each sub-network's own Top-200 and the global Top-200, i.e. the fraction of global Top-200 papers that also appear in the sub-network's Top-200. The results reveal a pronounced gradient, ranging from 89.5\% (Analytical Models) to 29.0\% (Emerging
Approaches). Categories that form the backbone of the field, namely Analytical Models~(1) at 89.5\%, Numerical Methods~(2) at 79.5\%, and European options at 78.5\%, exhibit high overlap, confirming that these categories constitute the shared intellectual infrastructure of option pricing research. Their citation priorities closely mirror the global consensus because nearly every sub-field builds upon the same foundational works in these areas.

At the other end of the spectrum, Machine Learning~(6) at 37.5\%, Market Imperfections~(4) at 44.0\%, and Emerging Approaches~(8) at 29.0\% show that more than half of these communities' most-cited works differ from the global Top-200. This reflects the fact that these sub-communities have distinct research agendas requiring different bodies of knowledge, and because they constitute a small fraction of the overall literature, their internal citation priorities are diluted in the global ranking. The contrast between $n_{\text{global}}$ and $n_{\text{local}}$ in Table~\ref{tab:overlap} illustrates this point: Machine Learning~(6) contributes only 2 papers to the global Top-200 but 72 to its own sub-network Top-200; Market Imperfections~(4) rises from 6 to 73; Behavioral~(7) from 16 to 70. These categories possess substantial bodies of internally important literature that are largely absent from the global ranking. Label-enhanced sub-network analysis makes these intrinsic patterns visible by allowing each sub-community's citation behaviour to be examined on its own terms, free from the dominance of larger categories.

\begin{table}[htbp]
\centering
\caption{Sub-network divergence from the global citation structure (Top-200). $n_{\text{global}}$ and $n_{\text{local}}$ are the number of papers belonging to each category in the global and sub-network Top-200, respectively. $\bar{r}_{\text{global}}$ and $\bar{r}_{\text{local}}$ are the mean ranks of those papers in their respective Top-200 pools (lower = more important). $\Delta r = \bar{r}_{\text{global}} - \bar{r}_{\text{local}}$: positive values indicate the category's papers are ranked higher in its own sub-network than globally. Overlap rate measures the fraction of the global Top-200 also appearing in the sub-network's Top-200.}
\label{tab:overlap}
\footnotesize
\begin{tabular}{llcccccr}
\toprule
\textbf{Dim} & \textbf{Category}
  & $\boldsymbol{n_{\text{global}}}$
  & $\boldsymbol{n_{\text{local}}}$
  & $\boldsymbol{\bar{r}_{\text{global}}}$
  & $\boldsymbol{\bar{r}_{\text{local}}}$
  & $\boldsymbol{\Delta r}$
  & \textbf{Overlap} \\
\midrule
\multirow{4}{*}{2}
& Stocks            &  64 &  83 &  87.0 &  99.7 & $-12.7$  & 69.0\% \\
& Indices           &  34 &  75 &  90.9 &  97.8 &  $-6.9$  & 49.0\% \\
& Currencies        &  12 &  44 &  90.9 &  84.9 &  $+6.1$  & 53.0\% \\
& Interest Rates    &  22 &  57 & 111.2 &  93.5 & $+17.7$  & 50.0\% \\
\midrule
\multirow{3}{*}{3}
& European          &  58 &  76 & 103.0 & 108.8 &  $-5.8$  & 78.5\% \\
& American          &  37 &  99 & 109.1 & 109.2 &  $-0.2$  & 61.0\% \\
& Exotic            &  36 &  87 & 121.9 & 109.3 & $+12.6$  & 66.0\% \\
\midrule
\multirow{8}{*}{4}
& Analytical (1)    & 163 & 173 & 100.2 & 101.9 &  $-1.8$  & 89.5\% \\
& Numerical (2)     &  93 & 117 & 110.0 & 109.2 &  $+0.7$  & 79.5\% \\
& Multi-Factor (3)  &  40 &  65 & 113.0 & 104.2 &  $+8.8$  & 73.0\% \\
& Mkt Imperf.\ (4)  &   6 &  73 & 124.2 &  99.9 & $+24.3$  & 44.0\% \\
& Calibration (5)   &  70 & 122 &  96.4 & 106.3 &  $-9.9$  & 60.0\% \\
& ML (6)            &   2 &  72 &  71.5 &  99.0 & $-27.5$  & 37.5\% \\
& Behavioral (7)    &  16 &  70 &  69.1 & 105.5 & $-36.3$  & 48.0\% \\
& Emerging (8)      &   0 &  30 &   --- &  78.3 &     --- & 29.0\% \\
\bottomrule
\end{tabular}
\end{table}

\subsubsection{Cross-Category Citation Preferences}

The overlap analysis above examines whether sub-networks identify different sets of important papers. We now ask a complementary question: holding the set of important papers fixed, how does each sub-community re-order them? This isolates the effect of citation preference from that of compositional change.

Concretely, we fix the global PageRank Top-200 as a common reference set and, for each sub-network $A$, re-rank these same 200 papers by sub-network $A$'s PageRank scores. Papers in the global Top-200 that receive no citations from sub-network $A$ are assigned the lowest rank. For each paper category $B$, we then compute the mean rank shift:
\begin{equation}
  \Delta R(A, B)
    = \bar{r}_{\text{global}}(B) - \bar{r}_{A}(B)
  \label{eq:rank_change}
\end{equation}
where $\bar{r}_{\text{global}}(B)$ is the mean rank of $B$-labeled papers in the original global ordering and $\bar{r}_{A}(B)$ is their mean rank after re-ordering by sub-network $A$. Because both quantities are computed over the same 200 papers, a positive $\Delta R$ purely reflects that sub-community $A$ assigns higher importance to $B$-papers than the global consensus does, without conflation from changes in pool composition.

\begin{figure}[htbp]
\centering
\includegraphics[width=\textwidth]{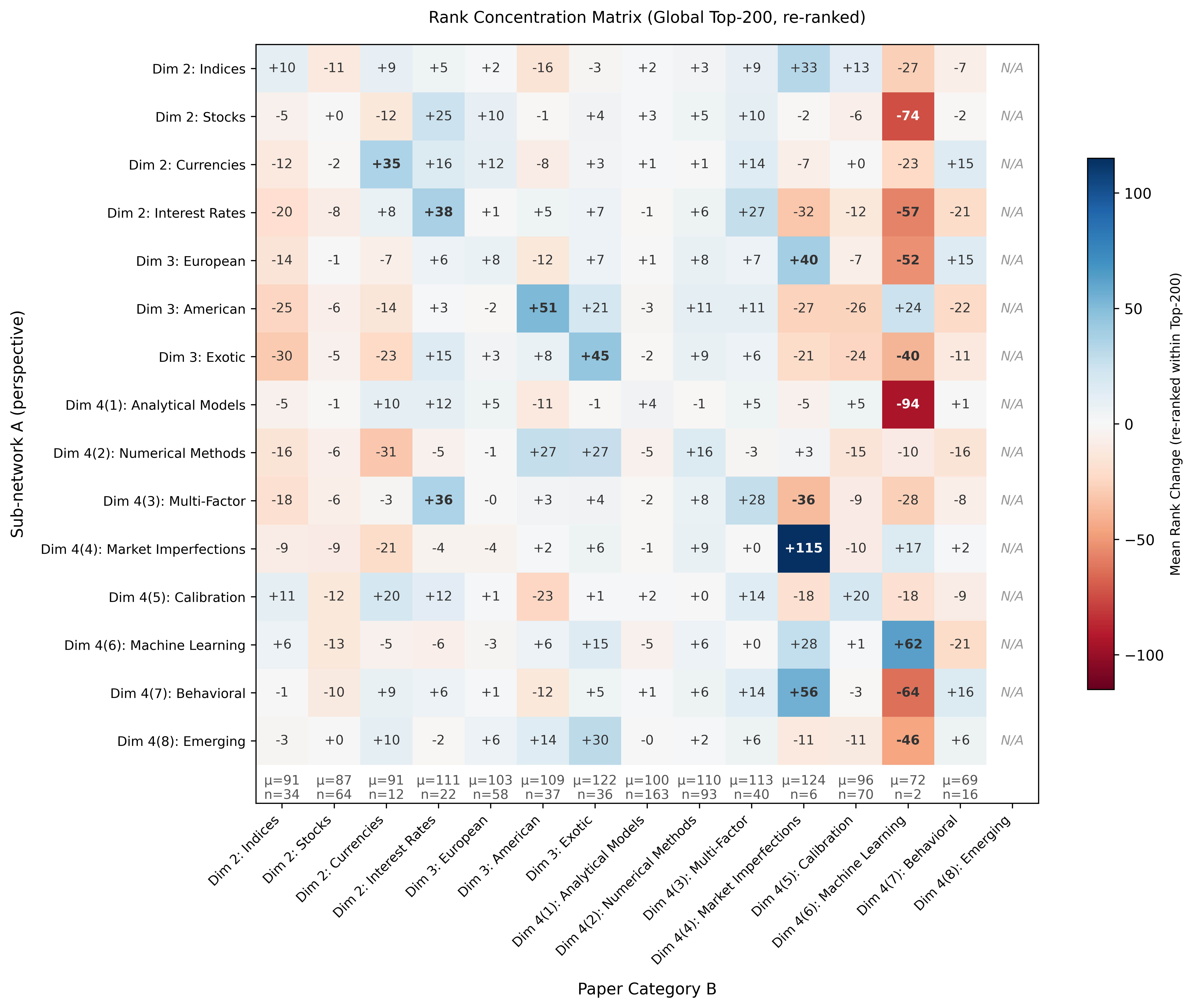}
\caption{Rank concentration matrix for the global PageRank Top-200. Each cell shows $\Delta R(A,B)$, the mean rank change of category $B$ papers when the Top-200 are re-ranked by sub-network $A$'s PageRank. Positive values (blue) indicate that $B$-papers are ranked higher from $A$'s perspective; negative values (red) indicate they are ranked lower. Bottom annotations show each category's mean global rank ($\mu$) and paper count ($n$) within the Top-200. Results are robust to Top-100 and Top-500 thresholds.}
\label{fig:rank_concentration}
\end{figure}

Fig.~\ref{fig:rank_concentration} presents the resulting matrix.
The findings are robust under Top-100 and Top-300 thresholds. We
highlight five structural patterns.
Several structural
patterns emerge.

\bmhead{Column-level patterns: universal vs.\ polarising categories}
The Analytical Models~(1) column exhibits near-zero $\Delta R$ values across all rows ($-5$ to $+5$), confirming that analytical works are equally valued by every sub-community regardless of research direction. Machine Learning~(6) presents the opposite pattern: its column is strongly negative from most traditional perspectives ($-94$ from Analytical Models, $-74$ from Stocks, $-64$ from Behavioral), indicating that classical sub-fields do not depend on ML literature. Notably, American options ($+24$) and Market Imperfections ($+17$) are exceptions, suggesting early engagement with ML through reinforcement learning for optimal exercise and data-driven friction modelling, respectively.

\bmhead{Diagonal patterns: self-concentration gradient}
The diagonal entries $\Delta R(A,A)$ reveal the degree to which each sub-community elevates its own literature when re-ranking the same papers. Analytical Models ($+4$) and Stocks ($+0$) show near-zero diagonals, confirming that their internal priorities coincide with the global consensus. Market Imperfections ($+115$), Machine Learning ($+62$), American options ($+51$), and Exotic options ($+45$) show strongly positive diagonals, indicating that these communities substantially re-order the shared Top-200 in favour of their own works. Combined with the low overlap rates in Table~\ref{tab:overlap}, this confirms that these sub-fields have developed distinctive citation ecosystems at both the set-composition and rank-ordering levels.

\bmhead{Off-diagonal patterns: inter-community dependencies}
Two notable off-diagonal relationships emerge. Interest Rates and Multi-Factor Models~(3) exhibit symmetric elevation: from the Interest Rates perspective, Multi-Factor papers rise by $+27$ positions; from the Multi-Factor perspective, Interest Rates papers rise by $+36$. This mutual dependence reflects the structural link between multi-factor stochastic frameworks and term-structure modelling. In contrast, Behavioral~(7) ranks Market Imperfections~(4) papers $+56$ positions higher, the second-largest off-diagonal value in the matrix, while the reverse is only $+2$. This asymmetry indicates that Market Imperfections literature is a prerequisite for Behavioral research but not vice versa.

\vspace{0.5em}
These inter-community dynamics are structural relationships that a global ranking cannot reveal. They become visible only when citation networks are decomposed by the content-level labels produced through LR-Robot's classification, without which it would not be possible to partition the network by methodological type, option type, or underlying asset.

\section{Conclusion}\label{sec.conclusion}

This study introduces LR-Robot, a human-in-the-loop LLM framework for systematic literature reviews that bridges two fundamental bottlenecks in large-scale academic literature analysis. On one hand, traditional expert-driven reviews are increasingly impractical given the exponential growth of publications, with manual screening and synthesis requiring months or years to complete. On the other hand, purely data-driven approaches, including unsupervised topic modeling and fully automated LLM pipelines, lack the domain-specific understanding needed for reliable content-level classification, as confirmed by our BERTopic experiments. LR-Robot resolves this tension through a clear division of labor: domain experts define classification taxonomies and prompt constraints that encode conceptual boundaries, while LLMs execute scalable classification under systematic human evaluation. The framework thus occupies a methodological niche that neither manual review nor fully automated methods can fill alone.

We validate LR-Robot on a corpus of 12,666 option pricing articles spanning five decades by introducing a four-dimensional classification scheme and systematically benchmarking multiple LLMs across tasks of increasing complexity. The results show that expert-designed prompt constraints significantly enhance classification performance. The framework demonstrates strong reliability on tasks with well-defined conceptual boundaries (Dims 1–3) and maintains stable, moderate performance on more complex classifications (Dim 4), even when relying solely on abstract-level information where category overlap is unavoidable. Error analysis indicates that most misclassifications arise from genuinely ambiguous cases rather than systematic model deficiencies, suggesting that performance limits are driven by the inherent difficulty of the task. At scale, the multi-dimensional labels generated by LR-Robot enable analyses that extend beyond traditional bibliometrics: temporal co-occurrence analysis uncovers trajectories of methodological co-evolution that are not captured by traditional methods and label-enriched citation networks reveal sub-community citation patterns. 

Several directions for future work emerge from this study. The current application relies on abstracts, which provide an efficient and scalable foundation for large-scale classification. However, abstracts may omit important methodological details and contextual nuances, which can limit the ability to distinguish between closely related research approaches. Extending the input to introduction part or even full-text could improve classification accuracy for more complex dimensions and enable a more detailed analysis of methodological evolution.
Additionally, cross-disciplinary validation in other fields characterised by high terminological overlap would help establish the generalisability of the expert-guided approach, while integration with dynamic knowledge graphs could support the identification of emerging research frontiers and predictive tracking of thematic shifts.\\

\backmatter







\noindent
\textbf{Acknowledgements} This work is funded by an Individual Funding Award from the School of Engineering Mathematics and Technology, University of Bristol, awarded to WW. WF's PhD research is supported by UKRI EPSRC Doctoral Training Partnership (EP/W524414/1).

\noindent
\textbf{Author Contributions} W.W. and J.Z. conceived and designed the framework, performed the experiments, and contributed to the main manuscript as well as to the discussion and revision of the content. Z.W. contributed to the design of the framework and conducted the LLM experiments. W.B. conducted the topic modeling experiments (BERTopic) and visualization.
\\

\noindent
\textbf{Data Availability} Data and code will be available upon request

\section*{Declarations}
\textbf{Conflict of interest} The authors declare no conflict of interest.

\bibliography{sn-bibliography}
\noindent

\newpage
\begin{appendices}

\section{BERTopic Modeling}\label{appendixbertopic}

\subsection{Evaluation Metrics and Hyperparameter Tuning}

To evaluate the performance of BERTopic models, we use topic quality (TQ), based on two established metrics, namely topic coherence (TC) and topic diversity (TD)~\citep{dieng-etal-2020-topic}. Specifically, topic coherence is measured using normalized pointwise mutual information (NPMI)~\citep{Lau2014MachineRT}, defined as
\[
\text{Topic Coherence} = \frac{1}{|T|} \sum_{t \in T} \frac{2}{N(N-1)} \sum_{i<j} \frac{\log \frac{P(w_i, w_j)}{P(w_i)P(w_j)}}{-\log P(w_i, w_j)},
\]
where $T$ denotes the set of topics, $N$ is the number of top words per topic, and $P(w_i, w_j)$ and $P(w_i)$ are the joint and marginal probabilities estimated from a reference corpus. Topic diversity~\citep{dieng-etal-2020-topic} is defined as
\[
\text{Topic Diversity} = \frac{\left|\bigcup_{t \in T} W_t\right|}{|T| \times N},
\]
where $W_t$ represents the set of top-$N$ words for topic $t$. The overall topic quality is computed as the product of coherence and diversity, i.e.,
\[
\text{Topic Quality} = \text{Topic Coherence} \times \text{Topic Diversity}.
\]

Using the above evaluation metrics, we adopt a random search--based strategy to explore key hyperparameters across the main components of BERTopic, where the optimal configuration is selected by maximizing overall topic quality. For UMAP, the tuned parameters include $n_{\text{neighbors}} \in \{15, 30, 50, \text{max\_nn}\}$ (where $\text{max\_nn} = \min(n_{\text{docs}} - 1, 100)$), $n_{\text{components}} \in \{5, 8, 10\}$, and $\text{min\_dist} \in \{0.0, 0.1\}$. For HDBSCAN, $\text{min\_cluster\_size} \in \{10, 15, 20, 30, 40\}$ and $\text{min\_samples} \in \{5, 10, 15, 20\}$ are considered. For the vectorization module, $\text{min\_df} \in \{1, 2, 3\}$ and $\text{ngram\_range} \in \{(1,1), (1,2)\}$ are tuned. In each trial, a parameter configuration is randomly sampled from the predefined search space, and the BERTopic model is fitted using precomputed document embeddings. A total of 24 trials were conducted, and the detailed results are reported in Table~\ref{table:hyperpar}.

\begin{table}[htbp]
\centering

\caption{Results of 24 random search trials for BERTopic hyperparameter tuning. The optimal configuration (Trial 8) achieves the best overall performance.}
\begin{tabular}{c c c c c c c c c}
\toprule
\textbf{Trial} & \textbf{\#Topics} & \textbf{n\_nb} & \textbf{n\_comp} & \textbf{min\_clust} & \textbf{min\_samp} & \textbf{TC} & \textbf{TD} & \textbf{TQ} \\
\midrule
0  & 97  & 15  & 5  & 15 & 10 & 0.0571 & 0.6505 & 0.0371 \\
1  & 172 & 15  & 10 & 10 & 5  & 0.0054 & 0.6669 & 0.0036 \\
2  & 45  & 30  & 10 & 40 & 10 & 0.1230 & 0.6978 & 0.0859 \\
3  & 103 & 30  & 8  & 10 & 10 & 0.1320 & 0.7068 & 0.0933 \\
4  & 73  & 50  & 8  & 15 & 15 & 0.0742 & 0.6658 & 0.0494 \\
5  & 48  & 100 & 5  & 20 & 15 & 0.1198 & 0.7125 & 0.0853 \\
6  & 50  & 15  & 8  & 40 & 15 & 0.1258 & 0.6620 & 0.0833 \\
7  & 119 & 30  & 10 & 10 & 10 & 0.0256 & 0.6555 & 0.0168 \\
8  & 55  & 30  & 5  & 20 & 20 & \textbf{0.1384} & 0.7200 & \textbf{0.0997} \\
9  & 74  & 30  & 8  & 15 & 15 & 0.0765 & 0.6676 & 0.0510 \\
10 & 69  & 30  & 10 & 15 & 20 & 0.1274 & 0.7232 & 0.0921 \\
11 & 103 & 30  & 10 & 10 & 10 & 0.1254 & 0.7068 & 0.0886 \\
12 & 64  & 100 & 8  & 15 & 15 & 0.1190 & 0.7328 & 0.0872 \\
13 & 7   & 100 & 10 & 15 & 15 & 0.0768 & 0.7714 & 0.0592 \\
14 & 5   & 50  & 10 & 40 & 20 & 0.0548 & \textbf{0.8200} & 0.0449 \\
15 & 145 & 30  & 10 & 10 & 5  & 0.0172 & 0.6669 & 0.0115 \\
16 & 40  & 30  & 10 & 40 & 5  & 0.1350 & 0.6725 & 0.0908 \\
17 & 34  & 100 & 10 & 40 & 5  & 0.0992 & 0.6412 & 0.0636 \\
18 & 6   & 50  & 10 & 10 & 15 & 0.0769 & 0.7833 & 0.0602 \\
19 & 33  & 100 & 5  & 40 & 10 & 0.0887 & 0.6606 & 0.0586 \\
20 & 67  & 50  & 10 & 15 & 15 & 0.0777 & 0.6552 & 0.0509 \\
21 & 80  & 50  & 8  & 10 & 15 & 0.0673 & 0.6575 & 0.0442 \\
22 & 88  & 15  & 5  & 10 & 20 & 0.0540 & 0.6705 & 0.0362 \\
23 & 39  & 30  & 10 & 40 & 10 & 0.1231 & 0.7179 & 0.0884 \\
\bottomrule
\end{tabular}

\label{table:hyperpar}
\end{table}

\subsection{Topic Modeling Results}

Table~\ref{table:bertopic} presents the topics identified by the BERTopic model, along with their keyword representations and corresponding descriptions.

The column ``Topic Keywords'' lists the most representative terms for each topic, which serve as the basis for the GPT-4.1-generated descriptions. Topic indices are assigned by the clustering algorithm and do not imply any inherent ordering or hierarchy.

The topic indexed as $-1$ corresponds to outliers, representing documents that cannot be reliably assigned to any coherent cluster during the density-based clustering stage. These documents typically exhibit heterogeneous content, overlapping themes, or weak semantic signals, making them difficult to group with other topics. As a result, the outlier category is relatively large compared to individual topics, reflecting the diversity and complexity of the option pricing literature.

\captionsetup{width=\textwidth}
\scriptsize
\begin{longtable}{cY Z c}

\caption{Topics identified by the champion BERTopic model, along with their keyword representations and GPT-4.1-generated descriptions.}
\label{table:bertopic} \\
\toprule
\textbf{Topic Index} & \textbf{Topic Keywords} & \textbf{GPT-4.1 Description} & \textbf{Count} \\
\midrule
\endfirsthead

\toprule
\textbf{Topic Index} & \textbf{Topic Keywords} & \textbf{GPT-4.1 Description} & \textbf{Count} \\
\midrule
\endhead

\bottomrule
\endfoot

-1 & \textbf{(Outliers)} option, volatility, pricing, model, options, price, prices, risk, market, option pricing & \textbf{(Outliers)} Option Pricing under Stochastic Volatility & 5298 \\
0 & stock, compensation, stock options, firms, executive, ceo, stock option, firm, employee, incentives & Executive Stock Option Compensation & 1408 \\
1 & neural, learning, network, neural network, networks, deep, neural networks, model, data, pricing & Neural Networks in Option Pricing & 551 \\
2 & real, project, investment, value, projects, real options, decision, option, carbon, real option & Real Options in Project Investment & 504 \\
3 & method, finite, numerical, scheme, order, american, difference, finite difference, differential, methods & High-Order Numerical Schemes for Options & 436 \\
4 & returns, market, implied, stock, trading, options, volatility, index, implied volatility, information & Informed Options Trading Predictability & 341 \\
5 & oil, volatility, market, implied, implied volatility, markets, indices, uncertainty, returns, index & Crude Oil Implied Volatility Dynamics & 329 \\
6 & fractional, fractional brownian, brownian, rough, motion, brownian motion, hurst, model, volatility, stochastic & Fractional Brownian Motion Option Pricing & 224 \\
7 & barrier, barrier options, barrier option, double, double barrier, options, barriers, option, pricing, monitored & Double Barrier Option Pricing Models & 210 \\
8 & parallel, gpu, computing, implementation, performance, cpu, architecture, hardware, pricing, monte carlo & GPU-Accelerated Option Pricing Algorithms & 194 \\
9 & jump, jumps, regime, switching, regime switching, model, stochastic, markov, diffusion, volatility & Regime-Switching Jump Diffusion Pricing & 182 \\
10 & tree, lattice, binomial, trinomial, binomial tree, pricing, trees, model, option, convergence & Binomial Tree Option Pricing Methods & 181 \\
11 & volatility, forecasting, implied, implied volatility, forecasts, realized, forecast, index, realized volatility, volatility forecasting & Implied Volatility in Forecasting & 180 \\
12 & fractional, time fractional, fractional black, black, method, order, time, numerical, fractional derivative, black scholes & Time-Fractional Black-Scholes Methods & 155 \\
13 & fuzzy, numbers, option, pricing, option pricing, model, price, fuzzy set, european, pricing model & Fuzzy Option Pricing Models & 121 \\
14 & credit, default, spreads, cds, credit default, credit risk, risk, credit spreads, debt, default risk & Credit Derivatives Valuation and Volatility & 120 \\
15 & hedging, problem, transaction costs, transaction, martingale, superhedging, costs, optimal, super, duality & Hedging Options with Transaction Costs & 119 \\
16 & rate, bond, rates, term structure, term, model, structure, short rate, factor, stochastic & Term Structure of Interest Rates & 108 \\
17 & clinical, health, received, research, patients, dr, fees, study, medical, pharmaceuticals & Medical Research Funding Disclosures & 108 \\
18 & perpetual, american, stopping, optimal, perpetual american, exercise, optimal stopping, american options, american option, boundary & Perpetual American Option Stopping & 103 \\
19 & futures, commodity, prices, convenience, futures prices, futures price, convenience yield, price, oil, model & Commodity Futures Pricing Models & 95 \\
20 & density, risk neutral, neutral, densities, neutral density, estimation, nonparametric, neutral densities, risk, option prices & Nonparametric Risk-Neutral Density Estimation & 92 \\
21 & monte, monte carlo, carlo, quasi monte, reduction, variance, qmc, quasi, variance reduction, method & Monte Carlo Methods in Finance & 91 \\
22 & garch, garch models, garch model, models, garch option, model, option pricing, pricing, option, innovations & GARCH Models in Option Pricing & 85 \\
23 & currency, foreign, exchange, exchange rate, currency options, foreign exchange, fx, rate, currency option, rates & Foreign Currency Option Pricing Models & 83 \\
24 & inverse, inverse problem, regularization, problem, local, local volatility, volatility, volatility function, ill, tikhonov & Local Volatility Surface Calibration & 82 \\
25 & implied volatility, asymptotics, asymptotic, volatility, implied, small, sabr, maturity, large, expansion & Small-Time Implied Volatility Asymptotics & 78 \\
26 & radial, radial basis, rbf, basis, basis function, basis functions, method, functions, function, numerical & Radial Basis Functions for Option Pricing & 78 \\
27 & uncertain, stock model, pricing formulas, formulas, lookback, uncertain stock, pricing, stock, uncertain financial, option & Uncertain Models for Option Pricing & 77 \\
28 & insurance, life insurance, life, mortality, guaranteed, linked, death, contracts, insurance contracts, guarantees & Life Insurance Valuation and Risk & 72 \\
29 & monte carlo, monte, carlo, american, squares, american options, squares monte, regression, method, simulation & Monte Carlo American Option Pricing & 71 \\
30 & vulnerable, default, counterparty, vulnerable options, credit, risk, default risk, credit risk, pricing vulnerable, options & Vulnerable Options and Credit Risk & 65 \\
31 & asian, asian options, asian option, geometric, arithmetic, average, geometric asian, options, price, arithmetic asian & Asian Options Pricing Models & 58 \\
32 & american, boundary, exercise, american options, exercise boundary, american option, optimal exercise, approximation, early, early exercise & American Put Option Exercise Boundary & 58 \\
33 & processes, stochastic, black scholes, scholes, martingale, black, process, vy, equation, option pricing & Stochastic Processes in Option Pricing & 56 \\
34 & distribution, kurtosis, log, distributions, gamma, model, density, pricing, motion, option pricing & Variance Gamma Option Pricing & 49 \\
35 & bank, deposit, banks, deposit insurance, capital, insurance, risk, loan, government, loans & Deposit Insurance and Capital Regulation & 48 \\
36 & supply, supply chain, chain, retailer, contract, option contract, supplier, demand, financing, procurement & Supply Chain Option Contracts & 41 \\
37 & wavelet, wavelets, method, wavelet based, discretization, legendre, fractional, equations, matrices, scheme & Wavelet Methods for Option Pricing & 39 \\
38 & stochastic, stochastic volatility, volatility, heston, volatility model, heston model, model, expansion, multiscale, approximation & Multiscale Stochastic Volatility Option Pricing & 38 \\
39 & mortgage, prepayment, mortgages, loan, house, value, rate, borrower, house price, mortgage backed & Mortgage Default and Option Pricing & 38 \\
40 & barrier, monte carlo, monte, carlo, barrier options, simulation, algorithm, pricing barrier, options, method & Efficient Monte Carlo Barrier Option Pricing & 38 \\
41 & basket, basket options, approximation, basket option, pricing basket, log, moment matching, matching, closed, closed form & Closed-Form Basket Option Pricing & 31 \\
42 & warrants, warrant, stock, pricing, warrant prices, structured, market, warrant pricing, stocks, issued & Warrant Pricing and Valuation Models & 31 \\
43 & quantum, amplitude, algorithms, algorithm, computing, circuit, monte carlo, monte, carlo, classical & Quantum Algorithms for Option Pricing & 30 \\
44 & fourier, cos, cos method, method, cosine, fourier cosine, fourier transform, transform, fast, cosine series & Fourier-Cosine Option Pricing Methods & 28 \\
45 & static, hedging, static hedging, hedge, barrier, hedges, barrier options, static hedges, options, portfolio & Static Hedging of Barrier Options & 28 \\
46 & bounds, upper, good deal, lower, prices, bound, bounds option, deal, upper bounds, lower bounds & Option Pricing Bounds in Incomplete Markets & 28 \\
47 & weather, temperature, rainfall, weather derivatives, derivatives, temperatures, cooling, daily, process, degree & Temperature Derivatives in Agriculture/Energy & 26 \\
48 & implied volatility, implied, volatility, formula, black, black scholes, scholes, approximation, method, scholes formula & Implied Volatility Approximation Methods & 25 \\
49 & lie, symmetry, symmetries, equation, group, invariant, classification, solutions, symmetry analysis, black & Lie Symmetry Analysis in Finance & 25 \\
50 & currency, volatility, iv, exchange, implied, usd, implied volatility, exchange rate, rate volatility, month & Currency Options Volatility Modeling & 23 \\
51 & reflected, bsdes, backward stochastic, stochastic differential, backward, reflected bsdes, stochastic, differential, uniqueness, solution & Reflected G-BSDEs and Applications & 23 \\
52 & wave, solutions, ivancevic, soliton, ivancevic option, nonlinear, pricing model, rogue, dark, option pricing & Ivancevic Option Pricing Waves & 22 \\
53 & copula, copulas, dependence, bivariate, spread, dynamic, spread option, gaussian, model, pricing & Dynamic Copula Bivariate Option Pricing & 21 \\
54 & volatility, implied, index, implied volatility, 500, models, model, stochastic, volatility function, data & Stochastic Volatility in Option Pricing & 21 \\
\end{longtable}

\section{Prompt for classification}\label{appendix1}
\subsection{Prompt for option pricing model classification:}
\label{subsection: appendix_pricing_model_classification}
\begin{quote}
Please clarify whether the abstract discusses developing or comparing pricing models or volatility models. I need a response that uses only the options listed below: [Yes, No]. What is your answer? Your answer should consist solely of the item from the list and nothing else. Your answer should also follow the constraints below:

\begin{itemize}
\item You should answer No if the abstract primarily focuses on the application of option pricing, rather than the development or comparison of option pricing models themselves.
\item You should answer Yes if the abstract focuses on methods of solving the existing option pricing or volatility model.
\item You should answer No if the abstract is about real estate investment or real option.
\item You should answer No if the abstract is purely about volatility and does not mention option pricing at all.
\item You should answer No if the abstract is purely about Greeks and risk management and does not mention option pricing at all.
\item You should answer No if the abstract is purely about hedging strategies and does not mention option pricing models at all.
\item You should answer No if the abstract describes a application of option pricing principles to a non traditional financial market.
\item You should answer No if the abstract is purely an empirical study testing the performance of existing, well-established option pricing models, without proposing any modifications or new solution methods.
\item You should answer No if the abstract focuses on market microstructure related to options, such as bid-ask spreads or trading volume, without discussing model development.
\item You should answer No if the abstract applies option pricing theory to model or predict bankruptcy or credit risk, without developing or comparing new option pricing models or solution methods.
\item You should answer No if the abstract primarily focuses on comparing or developing volatility models without a direct focus on option pricing models or their solution methods.
\item You should answer Yes if the abstract focuses on comparing different option pricing models, even if it involves an empirical study.
\item You should answer No if the provided text is a list of diverse paper topics from a proceedings or collection, rather than a single abstract focused on developing or comparing pricing/volatility models.
\item You should answer No if the abstract focuses on developing or comparing estimation methods for implied volatility surfaces, without directly developing or comparing option pricing models.
\item You should answer No if the abstract focuses on developing or analyzing numerical methods for solving PDE used in option pricing, without directly developing or comparing option pricing models.
\item You should answer No if the abstract applies option pricing theory to model or analyze insurance products, without developing or comparing new option pricing models or solution methods.
\item You should answer No if the abstract applies option pricing theory to model or analyze real options or investment opportunities, without developing or comparing new option pricing models or solution methods.
\item You should answer No if the abstract talks about cash-settled American-style options
\item You should answer No if the abstract talks about energy markets
\item You should answer No if the abstract talks about weather derivatives
\item You should answer No if the abstract talks about employee stock options
\item You should answer No if the abstract talks about vulnerable chained options
\item You should answer No if the abstract contains the phrase 'The proceedings contain'
\end{itemize}
\end{quote}

\subsection{Prompt for option underlying types classification:}
\label{subsection: appendix_underlying}
\begin{quote}
Task: Classify Underlying Asset Type. Classify the underlying asset type of options mentioned in the abstract. We have six questions for you to answer. For each question, please respond with only 'yes' or 'no' and nothing else.
\begin{itemize}[label={}]
\item Q1: Does this abstract specify Stocks as underlying assets?
\item Q2: Does this abstract specify Indexes as underlying assets?
\item Q3: Does this abstract specify Commodities as underlying assets?
\item Q4: Does this abstract specify Currencies as underlying assets?
\item Q5: Does this abstract specify Interest Rates as underlying assets?
\item Q6: Does this abstract specify Cryptocurrencies as underlying assets?
\end{itemize}
Please merge your responses to the final output as the following format \{Stocks: your response for Q1, Indexes: your response for Q2, Commodities: your response for Q3, Currencies: your response for Q4, Interest Rates: your response for Q5, Cryptocurrencies: your response for Q6\}.
\end{quote}

\subsection{Prompt for pricing model types classification:}
\label{subsection: appendix_pricing_model_types}
\begin{quote}

Class-Level Task: Classify this abstract of an academic paper into the option pricing methodology taxonomy. Please only assign up to all applicable class from the taxonomy. Use the exact subclass index 1-8 provided below and give me just a list in form of [class\_index; class\_index]. 
\begin{itemize}[label={}]

\item Taxonomy\_index and Toxonomy name:
\item 1	\quad Analytical Models
\item 2 \quad	 Numerical Methods
\item 3	\quad Multi-Factor and Hybrid Models
\item 4	 \quad Market Imperfections and Frictions
\item 5	\quad Calibration and Model Estimation
\item 6	\quad Machine Learning and Data-Driven Approaches
\item 7	\quad Behavioral and Alternative Paradigms
\item 8	\quad Emerging and Niche Approaches or Others(cannot find in the previous class)
\end{itemize}
\end{quote}

\begin{quote}
Subclass-Level Task: Classify this abstract of an academic paper into the option pricing methodology taxonomy. Please only assign up to all applicable subclass from the taxonomy. Use the exact subclass index [1.1, ...,8.3] provided below and give me just a list in form of [subclass\_index; subclass\_index]. The taxonomy index and toxonomy are as followings:

\begin{itemize}[label={}]
\item 1.1 \quad Analytical Models: Black-Scholes Extensions
\item 1.2 \quad Analytical Models: Stochastic Volatility Models
\item 1.3 \quad Analytical Models: Jump/Discontinuity Models
\item 1.4 \quad Analytical Models: Regime-Switching Models
\item 1.5 \quad Other Analytical Models
\item 2.1 \quad Numerical Methods: PDE/PIDE Solvers
\item 2.2 \quad Numerical Methods: Monte Carlo Simulation
\item 2.3 \quad Numerical Methods: Lattice/Tree Methods
\item 2.4 \quad Numerical Methods: Transform Methods
\item 2.5 \quad Other Numerical Methods
\item 3.1 \quad Multi-Factor and Hybrid Models: Stochastic interest rates/term structure of interest rates
\item 3.2 \quad Multi-Factor and Hybrid Models: Stochastic dividends
\item 3.3 \quad Multi-Factor and Hybrid Models: Multi-asset correlation
\item 3.4 \quad Multi-Factor and Hybrid Models: Hybrid local-stochastic volatility
\item 3.5 \quad Other Multi-Factor and Hybrid Models
\item 4.1 \quad Market Imperfections and Frictions: Transaction costs
\item 4.2 \quad Market Imperfections and Frictions: Illiquidity/funding costs
\item 4.3 \quad Market Imperfections and Frictions: Taxes/regulation
\item 4.4 \quad Other Market Imperfections
\item 5.1 \quad Calibration and Model Estimation: Implied volatility fitting
\item 5.2 \quad Calibration and Model Estimation: Density recovery
\item 5.3 \quad Calibration and Model Estimation: Statistical calibration
\item 5.4 \quad Other Calibration and Model Estimation
\item 6.1 \quad Machine Learning and Data-Driven Approaches: Neural PDE solvers/Deep learning for pricing prediction
\item 6.2 \quad Machine Learning and Data-Driven Approaches: Reinforcement Learning for optimal exercise
\item 6.3 \quad Machine Learning and Data-Driven Approaches: ML for calibration
\item 6.4 \quad Other Machine Learning and Data-Driven Approaches
\item 7.1 \quad Behavioral and Alternative Paradigms: Utility-based pricing
\item 7.2 \quad Behavioral and Alternative Paradigms: Behavioral biases
\item 7.3 \quad Behavioral and Alternative Paradigms: Ambiguity aversion
\item 7.4 \quad Other Behavioral and Alternative Paradigms
\item 8.1 \quad Emerging and Niche Approaches: Quantum computing
\item 8.2 \quad Emerging and Niche Approaches: ESG-adjusted models
\item 8.3 \quad Others (cannot find in the previous class)
\end{itemize}

\end{quote}

\end{appendices}

\newpage

\end{document}